\documentclass[runningheads,english]{llncs}

\usepackage{placeins}

\usepackage{graphicx}
\usepackage{xparse}   
\usepackage{ifluatex} 
\usepackage{ifvtex} 
\usepackage{iftex}   

\usepackage{amsmath,amssymb,amsfonts}
\usepackage{paralist}
\usepackage{booktabs}
 \usepackage{multirow}
\usepackage{fancyvrb}
\usepackage{cite}
\usepackage[T1]{fontenc}
\usepackage{amsmath}
\usepackage[most]{tcolorbox}
\usepackage{listings, color}

\usepackage{hyperref}
\hypersetup{
    colorlinks=true,
    linkcolor=blue,
    filecolor=magenta,      
    urlcolor=cyan,
}

\usepackage{csquotes}
\MakeOuterQuote"

\newcommand{\nop}[1]{{}} 

\usepackage{hyperref}
\newcommand{\webref}[2]{\href{#1}{#2}}

\usepackage{geometry}
\usepackage{tikz}
\usepackage{pgfplots}

\usetikzlibrary{positioning}
\usepackage{pgfplots}
\usepackage{csvsimple}
\usetikzlibrary{calc} 
\usepackage{lmodern}
\usepackage{graphicx}

\usepackage{todonotes}

\usepackage{tikz}
\usepackage{ulem}
  {\small \em \tt ccc\begin{table} aaa bbb}%
  {\end{table} ddd}

\newcommand{\xds}{\ensuremath{.*}}
\newcommand{\xdss}{\ensuremath{.\!*\!*}}
\renewcommand{\xdss}{\ensuremath{..}}
\newcommand{\xarrs}{\ensuremath{[*]}}
\newcommand{\fb}{\,?\,}
  
\newcommand{\simple}{simple}

\newcommand{\complex}{complex}
\newcommand{\Complex}{Complex}
\newcommand{\json}{JSON}
\newcommand{\kw}[1]{\text{\bf #1}}
\renewcommand{\kw}[1]{\ensuremath{\mathtt{#1}}}
\newcommand{\qkw}[1]{\ensuremath{\mathtt{\QQ{#1}\QQ}}}
\newcommand{\key}[1]{\ensuremath{\mathit{#1}}}
\newcommand{\qkey}[1]{\ensuremath{\mathit{\QQ{#1}\QQ}}}
\newcommand{\akey}[1]{\ensuremath{\mathsf{#1}}}
\newcommand{\qakey}[1]{\ensuremath{\mathsf{\QQ{#1}\QQ}}}

\newcommand{\xnot}{\kw{not}}
\newcommand{\qnot}{\qkw{not}}
\newcommand{\xtrue}{\kw{true}}

\newcommand{\xfalse}{\kw{false}}

\newcommand{\xnull}{\kw{null}}

\newcommand{\xone}{\kw{oneOf}}
\newcommand{\qone}{\qkw{oneOf}}
\newcommand{\xany}{\kw{anyOf}}

\newcommand{\xall}{\kw{allOf}}
\newcommand{\qall}{\qkw{allOf}}

\newcommand{\xreq}{\kw{required}}
\newcommand{\qreq}{\qkw{required}}
\newcommand{\xexcl}{\kw{excluded}}
\newcommand{\qexcl}{\qkw{excluded}}
\newcommand{\xpattReq}{\kw{patternRequired}}

\newcommand{\xtype}{\kw{type}}
\newcommand{\qtype}{\qkw{type}}

\newcommand{\xprops}{\kw{properties}}
\newcommand{\qprops}{\qkw{properties}}

\newcommand{\xpattProps}{\kw{patternProperties}}
\newcommand{\qpattProps}{\qkw{patternProperties}}
\newcommand{\xminP}{\kw{minProperties}}

\newcommand{\xthen}{\kw{then}}

\newcommand{\xif}{\kw{if}}

\newcommand{\xelse}{\kw{else}}

\newcommand{\xite}{\xif-\xthen-\xelse}

\newcommand{\xaddProps}{\kw{additionalProperties}}

\newcommand{\xpatt}{\kw{pattern}}

\newcommand{\xcont}{\kw{contains}}
\newcommand{\qcont}{\qkw{contains}}

\newcommand{\xminIt}{\kw{minItems}}
\newcommand{\qminIt}{\qkw{minItems}}
\newcommand{\xmaxIt}{\kw{maxItems}}

\newcommand{\xit}{\kw{items}}
\newcommand{\qit}{\qkw{items}}

\newcommand{\xdeps}{\kw{dependencies}}

\newcommand{\xenum}{\kw{enum}}

\newcommand{\xconst}{\kw{const}}

\newcommand{\xdref}{\kw{\$ref}}
\newcommand{\qdref}{\qkw{\$ref}}
\newcommand{\xderef}{\kw{\$eref}}
\newcommand{\qderef}{\qkw{\$eref}}

\newcommand{\xdcomm}{\kw{\$comment}}

\newcommand{\xdescr}{\kw{description}}
\newcommand{\qdescr}{\qkw{description}}

\newcommand{\xdefs}{\kw{definitions}}

\newcommand{\xtitle}{\kw{title}}
\newcommand{\qtitle}{\qkw{title}}
\newcommand{\xname}{\kw{name}}

\newcommand{\xtags}{\kw{tags}}

\newcommand{\xcomment}{\kw{comment}}

\newcommand{\xformat}{\kw{format}}

\newcommand{\xobject}{\akey{object}}
\newcommand{\qobject}{\qakey{object}}

\newcommand{\xstr}{\akey{string}}
\newcommand{\qstr}{\qakey{string}}

\newcommand{\qarray}{\qakey{array}}

\newcommand{\xdschema}{\kw{\$schema}}

\newcommand{\eqdef}{\ensuremath{\stackrel{\vartriangle}{=}}}

\newcommand{\gcomment}[1]{}
\newcommand{\hide}[1]{}
\newcommand{\code}[1]{}

\newcommand{\Iff}{\Leftrightarrow}

\newcommand{\VerThree}{Draft-03}
\newcommand{\VerFour}{Draft-04}
\newcommand{\VerFive}{Draft-05}
\newcommand{\VerSix}{Draft-06}
\newcommand{\VerSeven}{Draft-07}
\newcommand{\VerEight}{Draft 2019-09}

\newcommand{\JS}{JSON Schema}

\newcommand{\custcom}[2]{\marginpar{\tiny #1: {#2}}}

\newcommand{\GG}[1]{\custcom{giorgio}{#1}}

\newcommand{\AB}[1]{\custcom{amine}{#1}}

\newcommand{\M}{\ |\ }

\newlength{\NL}
\newcommand{\NN}{\ensuremath{\ \hat{}\ }}
\newcommand{\QQ}{\textnormal{\textquotedbl}}

\newcommand{\jsonsch}{JSON Schema} 

\renewcommand{\custcom}[2]{}

\begin{document}
%
\title{An Empirical Study on the ``Usage of Not'' \break in Real-World JSON Schema Documents \break \large (Long Version)}
\titlerunning{``Usage of Not'' in Real-World JSON Schema Documents}
%
\author{Mohamed-Amine Baazizi\inst{1}
\and Dario Colazzo\inst{2}
\and Giorgio Ghelli\inst{3}
\and Carlo Sartiani\inst{4}
\and Stefanie Scherzinger\inst{5}}
\authorrunning{Baazizi, Colazzo, Ghelli, Sartiani and Scherzinger} 
%
\institute{Sorbonne Universit\'e, LIP6 UMR 7606, France
\email{baazizi@ia.lip6.fr}
\and 
Universit\'e Paris-Dauphine, PSL Research University, France
\email{dario.colazzo@dauphine.fr}
\and
Dipartimento di Informatica, Universit\`a di Pisa, Italy
\email{ghelli@di.unipi.it}
\and
DIMIE, Universit\`a della Basilicata, Italy
\email{carlo.sartiani@unibas.it}
\and
University Passau, Passau, Germany 
\email{stefanie.scherzinger@uni-passau.de}
}
\maketitle              
\begin{abstract}
In this paper, we study the usage of negation in JSON Schema data modelling.
%
Negation is a logical operator that is rarely present in type systems
and schema description languages, since it complicates decision problems.
As a consequence, many software tools, but also formal frameworks
for working with JSON Schema, do not fully support negation.
As of today, the question whether covering negation is practically relevant, or 
a mainly theoretical exercise (albeit challenging), is open.
This motivates us to study whether negation
is really used in practice, for which aims, and whether it
could be --- in principle --- replaced by simpler operators.
We have collected the most diverse corpus of JSON Schema documents analyzed so far,
based on a crawl of 90k open source schemas hosted on GitHub.
We perform a systematic analysis, quantify
usage patterns of negation, and also qualitatively analyze schemas.
We show that negation is indeed used, following a stable set of patterns,
with the potential to mature into \emph{design patterns}.

\end{abstract}

\newcommand{\refschema}[1]{(\webref{https://github.com/sdbs-uni-p/json-schema-corpus/blob/main/json_schema_corpus/pp_#1.json}{schema \ #1 \ in our corpus})}

\section{Introduction}

Negation is a logical operator that is rarely present in type systems
and schema description languages, since it complicates decision problems.
{\json} Schema introduced negation, together with the And, Or, and XOr
operators, in {\VerFour} (2013). We study here whether negation
is really used by {\json} Schema users, for which aims, and whether it
could be in principle replaced by simpler operators.

For this aim, we collected a big set of {\json} Schema files out of GitHub
and started a systematic analysis, based on numerical quantification of 
usage patterns of negation and direct analysis
of schemas with negation.

We discovered that negation is indeed used by {\json} Schema users, according
to a set of usage patterns that we are going to describe. We also
discovered some peculiar uses, that we will describe later.

\section{Method}

\subsection{Schema collection and duplicate elimination}

In order to collect the schemas to analyse, we proceeded as follows.

\GG{Stefanie, can you update with July 2020 data?}\AB{updated}

We used the cloud-service Google BigQuery to identify all  {\JS}   documents with an open source license (excluding the specification of  {\JS}   drafts) on GitHub.
Our query was executed in July 2020 and  identified 91,600 URLs.
Of these, we could successfully retrieve 85,600 
 (using \verb!wget!). 
 We searched this collection for all files ending in {\tt .json},
containing a property {\tt \$schema} that identifies it as a schema, but 
not being the {\JS} declaration itself.
Having discarded all files with invalid JSON syntax,
we are left with 82,000 files, 
which became 24,000 after the elimination of perfect copies
of the same schema.
To our knowledge, this is the largest collection of real-world {\JS} 
documents ever analyzed so far.
\GG{Actually, I found quite a number of copies of the
{\JS} schema}

\code{
--trivial duplicate elimination, producing 23683 distinct schemas
drop table if exists uniq;
create table uniq
as
(select doc->'schema_file' as sch, count(*) as c
from dist
group by doc->'schema_file');
select count(*) from uniq;
}

Analysing these files, we realized that, still, many of them were just different versions of essentially
the same file: among the 24,000 different schemas, we found for example 14 different versions of a schema with title ``A JSON Schema for Swagger 2.0 API.'' (in the original set of 82,000 schemas
these 14 versions were replicated over 1,147 files). 

The presence of almost-duplicates is a problem, for some different reasons.
A first reason is that duplication creates skews in our numeric statistics. This is not very important,
since these statistics are interesting as they indicate trends, but the actual numbers are not crucial, as will be discussed again in the paper.

The most important reason is the very practical fact that we stumbled many times across the same pattern
during direct exploration, which is very time-consuming, and moreover it was not easy to understand whether 
the identical repetition of a pattern indicated that many designers
were following the same approach or just the presence of many versions of the same schema.
\code{
--count how many different descriptions exist
drop table if exists descriptions;
create table descriptions as
(select sch,
	   concat(right(sch#>>'{id}',260),'---',
			  right(sch#>>'{"$id"}',260),'+++',
			  right(sch#>>'{title}',260),'---',
              right(sch#>>'{description}',260)) as description,
       length(sch::text) as len,
       c
from uniq);
drop table if exists gdescriptions;
create table gdescriptions as
(select description,
       count(*) as versions, sum(c) as copies, min(len), max(len), 
       round(avg(len)) as avg, round(stddev(len)) as stdd
from descriptions
group by description);

---different descriptions: 15295+4519 = 19814 su 23683
--
select count(*)
from gdescriptions
union 
select versions
from gdescriptions
where description = '---+++---'

--half of the description disappear when we move to distfiles2!!!!
select(
select count(distinct d.description) 
from gdescriptions d left join distfiles2 dd using(description)
where dd.description is null) as "lostDescr",
(
select count(distinct dd.description)
from distfiles2 dd ) as "dist2Descr"
from dual
}

A third problem is that duplication makes our dataset uselessly big, hence our analysis queries  
take more time to run.
For these reasons we tried to reduce this form of redundancy.

Hence, in order to reduce the number of similar files, we computed for each file how many times each of the
{\JS} keywords appears in that file and, for each equivalence class with respect to this measure,
we selected the longest file.
After this step we remained with 13,700 files.
\code{
select count(*) from distfiles2
}

We then performed a further reduction step, 
where we regarded all files having the same
 (id/\$id, title, description) triple as versions of the same schema, provided that at least one of these members
 was present in the schema, and we only selected 
 the longest of version out of each equivalence class. This brought the number down to 11.5K 
 files.
 
 At this point we started looking for \xnot\ assertions, but we soon realized that an important
 fraction of these have the form
 \begin{quote}
   \xnot : \{ \xdref : \qkey{reference} \}
\end{quote}

\xdref\ is an operator that refers to another JSON Schema object that may be in the same file 
or in a different file. The presence of \xdref\ pointers prevents us from analysing the intended
use of the \xnot\ keyword, hence we prepared a \emph{ref-expanded} version of our 82,000
files, where we tried to substitute any instance of \qdref :  \qkey{reference} with
an expanded instance \qderef : \emph{referred object}.
This is not always possible, since we were not always able to retrieve the file that is referenced,
in case of non-local references, and since some references were not correct, but we have been able
to expand the vast majority of references (the 83\%).
The referenced object may contain other references, and references may be recursive, but this
is not a problem 
since, for our aims, we only need to substitute every reference with the referenced 
object, with no further expansion.
Nevertheless, this one-level expansion has been sufficient to increase the average size of our schemas
by a factor of almost three.

\code{
--percentage of \$eref / \$ref
select(select count(*)
       from edftree
       where key = '$eref')*100/
	   (select count(*)
       from dftree
       where key = '$ref')
from dual

--expansion of set of schemas
select(select count(*)
       from edftree
       )*100/
	   (select count(*)
       from dftree)
from dual
}

We finally looked for instances of the \xnot\ keywords in these 11.5K 
 distinct schemas using a
combination of automated counting techniques, to get statistics about 
pattern frequencies, and direct exploration, where we looked at how programmers
use negation and what are they trying to express.
       
\subsection{Pattern quantification}

In order to quantify the relevance of each usage pattern, we proceed
as follows.

For each usage pattern we define a set of \emph{pattern queries}, that characterize the usage pattern, such as, for example, the query
$$\xdss\xnot.\xreq$$

that characterizes situations where we have a \xnot.\xreq\ path inside the schema, that is, where \xreq\ is a field of 
the object associated to a \xnot\ keyword.
For each query, we count the number of documents that contain at least one instance,
and the total number of instances in the document collection.

Patterns are expressed using the following language.

\[
\begin{array}{llllllllll}
p & ::= step \M step\ p  \\
step & ::= .key \M \xds \M  \xarrs \M  \xdss   \\
\mathit{filtered\ p} & ::= p {\fb} p
\end{array}
\]

Pattern matching is defined as in JSONPath \cite{JSONPath}: the step $.key$, applied to an object,
retrieves the member value whose name is $key$. 
The step $\xds$ retrieves all member values of an
object, $\xarrs$ retrieves all items of an array, and $\xdss$ is the reflexive and transitive closure of 
the union of $\xds$ and $\xarrs$, hence it navigates to all nodes of the JSON tree to which it
is applied.

Finally, we will use the conditional form $p1\fb p2$ to denote those nodes $n$ that are reached by a path $p1$ and
such that if we follow a path $p2$ starting from $n$ we will arrive to some node, so that, for example the query:
$$\xdss\xany \fb \xdss\xnot$$
indicate all subtrees reached by a path $\xdss\xany$ that include an instance of \xnot.

\subsection{Working with our data}

One important result of our research has been the construction of our dataset, that allows
researchers to analyse any kinds of aspects that relate to the real-world use of {\JS},
not only to use of negation.

This dataset is available to every researcher. In \cite{} we describe how to rebuild our
dataset on a local Postgres installation, how to check all numbers that we publish here, and how to build different queries related to other aspects of {\JS} usage.

\section{Description of the retrieved schemas}

\subsection{Versions of {\JS}}

{\JS} exists essentially in five versions: {\VerThree} of November 2010 \cite{Draft03}, 
{\VerFour} of February 2013 \cite{Draft04}, 
{\VerSix} of April 2917 \cite{Draft04}, 
{\VerSeven} of March 2018 \cite{Draft07}, 
{\VerEight} of September 2019 \cite{Draft08}; 
{\VerFive} was essentially a cleanup of {\VerFour} and used the same meta-schema,
while versions before {\VerThree} have been absorbed by that one.\GG{Add latest draft}\AB{done}

For each retrieved file, we analysed the \xdschema\ declarations that it contains in 
order to see the version of {\JS} to which it adheres. 
The distribution of the collected schemas across different versions of {\JS}
is reported in Table \ref{tab:drafts};
{\VerEight} (also known as Draft-08) is still quite new, hence is not really represented.
{\VerFour} is declared in the vast majority of the files, while 
{\VerSeven}, {\VerSix}, and the old {\VerThree}, are declared in comparable quantities.
It is worth adding that an analysis of files content showed us that the actual version
that a schema follows is often different from the version declared.

\begin{table}\label{tab:drafts}
\center
\small
\begin{tabular}{| l | r|  r  |}
\hline
Version	& Amount of files  & Percentage \\
\hline
draft-01	&	1	&	0.00	\\
draft-03	&	3106	&	3.78	\\
draft-04	&	64732	&	78.85	\\
draft-05	&	21	&	0.03	\\
draft-06	&	6787	&	8.27	\\
draft-07	&	2195	&	2.67	\\
no version specified	&	5252	&	6.40	\\
total	&	82094	&	100.00	\\
\hline
\end{tabular}
\caption{Distribution of files according to the version referred in their \$schema attribute}
\end{table}

\code{
select coalesce(f.draft,'total') as draft, '&',
       count(*) as count, '&',
	   round(count(*)*100.0/g.grandtotal,2) as percentage, '\\'
from fileschema f, 
     (select count(*) from fileschema) as g(grandtotal)
group by rollup(f.draft), g.grandtotal
order by draft;
}

For our study, it is worth recalling that the boolean operators \xnot, \xone, \xany, and \xall, have been
added with {\VerFour}. The possibility of using \xfalse\ and \xtrue\ wherever an assertion is 
expected has been introduced with {\VerSix}, together with the operator \xcont, while the \xite\ operators 
have been introduced with {\VerSeven}.

\subsection{Frequence of different keywords}\label{sec:frequence}

We counted the number of occurrences of keywords, and the number of files where each keyword
appears. The result is presented in table
\ref{tab:tableone}.

\code{
##### create general table of keywords
}

\begin{table}\label{tab:tableone}
\center
{\footnotesize
\begin{tabular}{| l | r | r | r | }
\hline
Keyword  & Occurrences   & Distinct files &  Draft-\\
\hline
Total	&	1347003	&	11508	&		\\
type	&	420780	&	11419	&	3	\\
description	&	222901	&	8022	&	3	\\
\$ref	&	179811	&	6288	&	3	\\
properties	&	87747	&	11110	&	3	\\
required	&	51476	&	8706	&	3	\\
oneOf	&	46786	&	2274	&	4	\\
enum	&	44383	&	5134	&	3	\\
items	&	44380	&	6345	&	3	\\
title	&	40903	&	6254	&	3	\\
additionalProperties	&	37111	&	5738	&	3	\\
id	&	22963	&	4794	&	4	\\
default	&	21298	&	2494	&	3	\\
pattern	&	18345	&	3016	&	3	\\
maxLength	&	15406	&	1472	&	3	\\
minLength	&	15324	&	1673	&	3	\\
\$schema	&	13681	&	11508	&	3	\\
minimum	&	11222	&	2015	&	3	\\
anyOf	&	8843	&	1504	&	4	\\
minItems	&	6830	&	1582	&	3	\\
maximum	&	5488	&	1275	&	3	\\
maxItems	&	4460	&	648	&	3	\\
uniqueItems	&	4390	&	1269	&	3	\\
definitions	&	4197	&	3750	&	4	\\
allOf	&	3813	&	1397	&	4	\\
patternProperties	&	2823	&	914	&	3	\\
examples	&	2765	&	356	&	3	\\
\$id	&	2116	&	748	&	6	\\
additionalItems	&	1276	&	401	&	3	\\
readOnly	&	1244	&	107	&	7	\\
not	&	787	&	298	&	4	\\
const	&	648	&	137	&	6	\\
exclusiveMinimum	&	496	&	143	&	3	\\
minProperties	&	465	&	230	&	4	\\
dependencies	&	416	&	221	&	3	\\
multipleOf	&	350	&	83	&	4	\\
\$comment	&	200	&	58	&	7	\\
exclusiveMaximum	&	196	&	60	&	3	\\
if	&	166	&	29	&	7	\\
then	&	166	&	29	&	7	\\
maxProperties	&	141	&	78	&	4	\\
deprecated	&	78	&	18	&	3	\\
propertyNames	&	54	&	26	&	6	\\
contains	&	45	&	10	&	6	\\
else	&	25	&	12	&	7	\\
writeOnly	&	6	&	5	&	7	\\
\$defs	&	2	&	2	&	8	\\
\hline
\end{tabular}
}
\caption{Occurrences of each keyword and number of files where it appears}
\end{table}

\code{
select coalesce(k.keyword,'Total'), '&',
       count(*) as c, '&',
	   case when keyword is not null then count(distinct d.line)
	   else (select count(*) from df2)
	   end, '&',
	   max(k.version), '\\'
from dftree d join keywords k on (d.key=k.keyword) 
where k.kclass in ('s', 'dollar')
group by rollup(k.keyword)
order by c desc;
}

\code{
--for ER paper
--listing ``user defined keyword''
-- total keys that are not keywords are 120,113 out of 1,487,366
-- the most common are:
value 6939 247
name 5328 983
rule 5085 342

select d.key, 
count(*) as count, count(distinct line) as files
from dftree d left outer join keywords k on (d.key = k.keyword)
where (ktype is null or ktype = 'ns') and --comment here
key not in ('format','readonly','example','schema') and --comment here
key not in ( '$','/')
and key not similar to '\_
and key not similar to '[0-9]
group by cube(d.key)
order by count desc

}

\gcomment{
{\JS} allows users to utilize non-standard keywords, and this is the reason for the fourth column:
we distinguish here standard keywords, marked as `s', and non-standard keywords, 
such as \xname, \xtags, \xcomment, which we divided 
into four groups that we will describe later.
We count here all standard keywords and only a chosen subset of the non-standard keywords, since this last
class is huge.}

A precise count of keyword occurrences presents some difficulties. Consider for example the following snippet 
\refschema{37687}:
{\small
\begin{verbatim}
"properties" : {
    "not": { "title": "...",
             "description": "...",
             "not": {"type": "null"}
    }, ...
}
\end{verbatim}
}
The outermost \xnot\ is not an occurrence of the \xnot\ keyword, but is just a property name, while the 
internal \xnot\
is an actual keyword occurrence. Of course, we cannot just avoid to count occurrences that are below \xprops, since
\xprops\ itself may be a property name, as in the following snippet of code \refschema{69248}
{\small
\begin{verbatim}
             "not": {
                    "anyOf": [
                        {
                            "properties": {
                                ...
                                "properties": {
                                    "not": {
                                        "properties": {
                                            "name": {
\end{verbatim}
}
Here, both occurrences of \xnot\ are occurrences of the operator, since the \xprops\ that surrounds the internal
\xnot\ is not a keyword but is a property name, since it is contained into a \xprops\ keyword.

In order to solve this problem, we rewrote all of our files and we prepended every occurrence of 
a member name that is in a position where a property name, rather than a keyword, is expected, with ``\_\_\_'',
and we did this through a complete parsing of the document, guided by {\JS} conventions.

Unfortunately, this is not yet enough. The problem is that programmers are allowed to extend {\JS} as they
like hence, when we meet a user-defined keyword, we have no way to know which fields to rename,
as happens in the following snippet \refschema{32451}.
{\small
\begin{verbatim}
{ "description": "...",
  "@errorMessages": { "not": "Invalid target: ..." },
  "not": {  "pattern": "..." }
  ...
}
\end{verbatim}
}
Here, ``@errorMessages'' is a user-defined keyword whose value is an object that describes
the error,  and not a {\JS}
assertion, hence the contained \xnot\ is not an occurrence of the {\JS} keyword but is just a member name
(the \xnot\ in the next line, on the contrary, is an occurrence of the keyword). Unfortunately, there are
other user-defined keyword whose value must be, on the contrary, interpreted as an assertion,
and we cannot know which user-defined keyword belongs to which category.

We tried two approaches. In the ``strict'' approach we renamed everything that was inside a user-defined
keyword, hence making it inaccessible by the analysis,
and in the ``lax'' approach we kept the content of any user-defined keyword with no renaming,
so that both instances of \xnot\ in the previous example were counted as keywords.
With the strict approach, some interesting usage patterns are lost, and keyword usage is under-estimated.
With the lax approach, we get some ``false positive'' as in the example above, hence some over-estimation.
One could also try a more sophisticated approach, where one tries to guess the best way of performing
renaming for each different user-defined keyword.
For our aims, we decided that the over-estimation of the lax approach was better than the under-estimation
of the strict one, and that a more sophisticated approach was not worth the extra effort, hence we adopted
the lax approach; the reader
should keep into account this detail while reading the tables.
The presence of user-defined keywords is far from negligible --- at a first analysis, we have 
more than 1 user-defined keyword occurrence for any 10 standard keyword occurrences.

\code{
-- counting the amount of user-defined keywords
with keyTypes as
(select  key,
        case when (key in (select keyword
						    from keywords
						    where kclass in ('s','dollar'))) then 'standard keyword'
			when key = '$'  then 'root object'
			when key like '\_\_\_
			when key similar to '[0-9]*' then 'array element'
            else 'user-defined keyword'
            end as typeOfKey,
			row_number () over (order by key) as rowNum
from dftree) 
select typeOfKey, count(*)
from keyTypes
--where mod(rowNum,1000) = 0
group by typeOfKey;
}

For our purposes, the most interesting outcome of this first analysis is the fact that \xnot\ appears in the 
3\% of all schemas, and it occupies the 30-th position in our table, out of 46 keywords analysed.
This indicates that \xnot, although it clearly belongs to the bottom half of {\JS} operators in terms
of usage, is indeed used by a non-negligible portion of {\JS} designers.

If we analyse the use of boolean operators in general, we notice that the most common boolean operator is, by far, 
\xone, whose occurrences are five times more common than \xany. 
We found that quite surprising, since the exclusive-disjunctive semantics of \xone\ is, in some sense, more complicated
than the purely disjunctive semantics of \xany. 
\custcom{}{\tiny would a survey help? we could survey {\JS} authors\dots}
\GG{Yes, a survey would add value to the paper}
A possible explanation of this preference 
is the fact that the description of a class as a \xone\ combination of a set of ``subclasses'' is a close relative of the 
exclusive-subclassing mechanism of object-oriented languages, which is a well-known jargon.
\xall\ is even less common. For any single occurrence of \xnot, we have 5 of \xall, 11 of \xany, and 59 of \xone.

The last boolean operator, \xite, is even less common than \xnot, but it has only been added with {\VerSeven}.

At this point, we know that \xnot\ appears 787 times in 298 different files.
This is a quite small fraction of the initial set of files, but the numbers are still
big enough to deserve a systematic study.


\section{Counting negated keyword}

\subsection{Counting \xnot.$k$ paths}

In order to analyse how users use negation, we first examined which are the {\JS} operators that are 
typically negated by a schema designer.
The result is in table \ref{tab:notargs}, where you find, for each keyword $k$, the frequency (number of occurrences)
of the \xnot.$k$ pattern, and the number of files where that pattern appears.

\code{##### create general table of keywords}

\begin{table}\label{tab:notargs}
\centering
\footnotesize
\begin{tabular}{| l  r r l r r |}
\hline 
 path	& occ.	 &	files  & path	& occ.	 &	files \\
\hline
not	&	787	&	298	&	not.\$eref	&	93	&	28	\\
\hline
not.*	&	840	&	289	&	not.\$eref.*	&	338	&	28	\\
not.required	&	240	&	84	&	not.\$eref.required	&	36	&	15	\\
not.items	&	126	&	27	&		&		&		\\
not.type	&	62	&	51	&	not.\$eref.type	&	51	&	20	\\
not.properties	&	71	&	47	&	not.\$eref.properties	&	40	&	18	\\
not.\$eref	&	93	&	28	&		&		&		\\
not.enum	&	61	&	52	&	not.\$eref.enum	&	12	&	8	\\
not.allOf	&	23	&	23	&	not.\$eref.allOf	&	38	&	5	\\
not.pattern	&	47	&	28	&		&		&		\\
not.anyOf	&	45	&	36	&	not.\$eref.anyOf	&	2	&	2	\\
not.description	&	4	&	4	&	not.\$eref.description	&	41	&	7	\\
not.title	&	2	&	2	&	not.\$eref.title	&	39	&	7	\\
	&		&		&	not.\$eref.\$schema	&	41	&	9	\\
not.\$fref	&	27	&	14	&		&		&		\\
not.oneOf	&	6	&	4	&	not.\$eref.oneOf	&	18	&	3	\\
not.additionalProperties	&	11	&	11	&	not.\$eref.additionalProperties	&	9	&	7	\\
not.patternProperties	&	15	&	15	&		&		&		\\
not.const	&	6	&	1	&		&		&		\\
	&		&		&	not.\$eref.definitions	&	3	&	3	\\
	&		&		&	not.\$eref.id	&	2	&	2	\\
	&		&		&	not.\$eref.dependencies	&	2	&	2	\\
	&		&		&	not.\$eref.not	&	2	&	2	\\
	&		&		&	not.\$eref.\$ref	&	2	&	2	\\
not.\$comment	&	1	&	1	&		&		&		\\
\hline
\multicolumn{3}{l}{Occurrences of \xnot\ without a not.* path}\\
\hline
\qnot : \{\ \}  & 16 & 8 &&& \\
\qnot : {\QQ}...\QQ  & 1 & 1 &&&\\
\hline
\end{tabular}
\caption{Occurrences of \xnot.$k$ paths}
\end{table}

At the left-hand-side of the table we analyse occurrences of \xnot.$k$ for different keywords.
As we anticipated in the introduction, \xnot\ is often followed by \xdref. In 93 cases out of
93+27 we have been able to expand that \qdref : \qkey{reference} into
a structure   \qderef : \emph{referenced schema}, and the result is analysed in 
the right-hand side of the table.

Observe that Table \ref{tab:notargs} indicates a total of 840 
occurrences of \xnot.$K$, but Table \ref{tab:tableone} has a total of 
787 occurrences of \xnot.
The two numbers differ since a single \xnot\ occurrence may generate
two or more \xnot.$K$ paths, or may generate none.
It generates more paths when its argument contains many fields, as in the following example.

{\small     
\begin{verbatim}
"not" : {"enum": ["generic-linux"], "type": "string"}
\end{verbatim}
}

Here, we say that \xnot\ has a \emph{\complex} argument since its argument contains more than one keyword,
and we will say that the two keywords \emph{co-occur} in the negated schema; 
otherwise we say that the schema is \emph{\simple}.

While most sub-schemas inside a schema are actually \complex,
most instances of \xnot\ have a \simple\ argument, that is, an argument with exactly one field.
We only found 55 instances out of 787
of \xnot\ with a \complex\ argument, most of them with just two keywords, but some with 
three or even four keywords in the negated schema, as in
\{\qtype:\qobject, \qtitle: \ldots, \qreq: \ldots, \qprops: \ldots\}.

\code{
-- 787 + (127-55= 72) = 859 - 17 = 842 - should be 840
-- distribution of valuelen
select (valuelen), count(*)
from treewithsiblings
where key = 'not' and valuelen > 1
group by valuelen

select sum (valuelen), count(*)
from treewithsiblings
where key = 'not' and valuelen > 1

--result:
0	16
1	716
2	41
3	11
4	3

--mostsubschemas are complex
select(
	select count(*) from treewithsiblings
	where sibnum > 1)*100/
	(
	select count(*) from treewithsiblings)
	from dual
}

On the other side, a \xnot\ may not correspond to any \xnot.$k$ pattern, when \xnot\ is followed by \{ \}.
We found 16 occurrences of \qnot : \{ \}, which are used to express the schema \xfalse\ that is not satisfied
by any instance. This use of \xnot\ is a consequence of the fact that the general use of \xfalse\ has only 
been introduced with {\VerSix}.

The situation is very different with \xderef: in this case we have that 93 occurrences
of \xnot.\xderef\ correspond to 338 occurrences of \xnot.\xderef.$k$. In this case, thanks
to the mediation of \xderef, the schema designer is implicitely applying negation to a 
\complex\ argument, with an average number of 3-4 members.

We sorted the table on the total number of  \xnot.$k$+\xnot.\xderef.$k$ occurrences, and
it is interesting to compare the weight of different keywords in the two table.

A surprising difference is the fact that the operator \xit\ never appears under \xnot.\xderef\
but it appears under \xnot\ with a high frequency, and we will discuss this later.

The most common argument of negation is clearly \xreq, which we will discuss later.
The operators \xtype\ and \xprops\ follow \xreq, with similar frequency.
Here, one notices that while \xnot.\xreq\ dominates the \xnot.$k$ case, 
the two most common cases of the 
\xnot.\xderef\ group are \xnot.\xderef.\xtype, whose value is \xobject\ in the 80\% of the cases,
and \xnot.\xderef.\xprops, which indicates that
\xnot.\xderef\ is mostly used to negate complex object definitions.

Another obvious difference is the much higher occurence of descriptive keywords, such as \xdescr, \xdschema, and \xtitle, inside referenced arguments with respect to the direct arguments of
\xnot. 

The last two lines of the table describe cases where \xnot\ is not followed by any keyword.
The case \qnot\ : \{\} has been already discussed. The case \qnot\ : \qkey{\dots} has been described in Section \ref{sec:frequence} and
is an artefact of a user-defined keyword ``@errorMessage'' that uses \xnot\ as the message name.

We now move to a detailed analysis of each keyword.

Keywords found below \xnot\ can be divided in three categories, that are called, in {\JS} jargon, \emph{assertions}, 
\emph{applicators}, and \emph{annotations}. 

Assertions include \xreq, \xenum, \xconst, \xpatt\ and \xtype, and indicate a test that is performed on the
corresponding instance.

Applicators include the boolean operators \xany, \xall, \xone, \xnot, the object operators \xprops, \xpattProps, \xaddProps,
the array operator \xit, and the reference operators \xdref, and they indicate a request to apply a different 
operator to the same instance or to a component of the current instance.

Annotation operators include \xtitle, \xdescr, and \xdcomm, they do not affect validation
but they indicate
an annotation that should be associated to the instance. Since we are mostly interested in validation, and since, moreover, annotations
are removed by the \xnot\ operator, we will ignore them.

In Section \ref{sec:assertions} we will describe the assertion keywords that are arguments of \xnot, and in 
Section \ref{sec:applicators} we will describe the applicators. In Section \ref{sec:contexts} we will look in the opposite
direction, and will analyse which applicators have \xnot\ as an argument. 

We now discuss assertions and applicators separately, but the separation is not total, since in a {\complex} schema
assertions and applicators may co-occur, as in the following example.
 
{\small     
\begin{verbatim}
"not" : { "type": "object",
          "properties" : {"a": { ... } }
}
\end{verbatim}
}

\subsection{Assertions: \xreq, \xenum, \xconst, \xpatt, \xtype}\label{sec:assertions}

\subsubsection{\xnot.\xreq: field exclusion}\label{sec:notreq} 

The most common operator that appears below \xnot\ is \xreq. 


The simplest, and most common case, of \xnot.\xreq\ path is when the argument of \xnot\ is \simple
(hence no other keyword co-occurs with \xreq),
and moreover the argument of \xnot\ is a one-string array, as in the following example.

{\small     
\begin{verbatim}
"not": { "required": ["DisplaceModules"] }
\end{verbatim}
}

Out of 240 occurrences of \xreq, 224 of them have this \simple\ shape, which hence constitutes alone the 27\% of all the 840 occurrences of \xnot.$k$ (Table \ref{tab:notreq}).

\begin{table}\label{tab:notreq}
\center
\begin{tabular}{| l | l |r|  r  |}
\hline
Simple/complex & excluded fields & occurrences & files \\
\hline
*	&	*	&	240	&	84	\\
simple	&	*	&	224	&	74	\\
complex	&	*	&	16	&	13	\\
*	&	onefield	&	173	&	56	\\
*	&	twofields	&	64	&	34	\\
*	&	manyfields	&	3	&	3	\\
simple	&	onefield	&	162	&	48	\\
simple	&	twofields	&	60	&	31	\\
complex	&	onefield	&	11	&	8	\\
complex	&	twofields	&	4	&	4	\\
simple	&	manyfields	&	2	&	2	\\
complex	&	manyfields	&	1	&	1	\\
\hline
\end{tabular}
\caption{The structure of arguments of \xnot\ that contain \xreq.}
\end{table}

\code{
-- lines on notarg use required
select coalesce(issimple,'*'), '&', 
       coalesce(howmany,'*'), '&', 
        count(*), '&', 
		count(distinct t.line) as numfiles, '\\',
	    array_agg(distinct sibnum) as lengths,
		array_agg(distinct valuelen) as lenArgs,
		array_agg(distinct t.line) as lines, 
        jsonb_agg(distinct t.sibkeys) as keys,
        jsonb_agg(distinct t.value) as args
from treewithsiblings t,
     ift (sibnum = 1, 'simple'::text, 'complex'::text) as issimple,
     ift (valuelen=1, 'onefield',
		   ift (valuelen=2, 'twofields','manyfields')) 
	     as howmany
where path like '
group by cube(issimple, howmany)
order by case when issimple is null and howmany is null 
               then 1
			   when howmany is null then 2
			   when issimple is null then 3
			   else 4 end,
      count desc;
}

While ``not required'' may sound like ``optional'', it actually means
that the object must violate the \qreq: ["DisplaceModules"]
specification, that is, that the field \QQ DisplaceModules\QQ\ must be absent.

\begin{remark}\label{rem:implication}.
One may imagine to add an operator \qexcl: \qkey{k}to {\JS}, 
that specifies that a field \key{k} cannot be present, with the idea that
\qnot: \{ \qreq: [\qkey{k}] \}
would then be the same as \qexcl: \qkey{k}, but we must be careful about the precise meanings
of {\JS} assertions.
By definition,  \qreq: [\qkey{k}] means:
\begin{quote}
if the instance is an object, then it contains
the field  \qkey{k}
\end{quote}
in other words, a string or an integer satisfies \qreq: [\qkey{k}].
Assume that we added an operator 
\qexcl: \qkey{k}
with the usual implicative meaning:
\begin{quote}
 if the instance is an object, then it does not contain
the field  \qkey{k}.
\end{quote}
In this case, we would have the following equivalences:

\textnormal{
\qnot: \{ \qreq: [\qkey{k}] \} $\ \Iff\ $
\{\qtype : \qobject, \qexcl: \qkey{k}\}
}

and 

\textnormal{
\qnot: \{ \qtype : \qobject, \qreq: [\qkey{k}] \} $\ \Iff\ $
\{ \qexcl: \qkey{k}\}
}

Of course, in a context that forces the instance to be an object, the two operators would be one the opposite of the
other, hence we would have the following equivalences:

\textnormal{
\{ \qtype : \qobject, \qnot: \{  \qreq: [\qkey{k}] \} \} \\
\ \ \ \ \ \ $\ \Iff\ $ \{ \qtype : \qobject, \{ \qexcl: \qkey{k}\}  \} \\
\{ \qtype : \qobject, \qnot: \{  \qexcl: [\qkey{k}] \} \} \\
\ \ \ \ \ \ $\ \Iff\ $ \{ \qtype : \qobject, \{ \xreq: \qkey{k}\}  \}
}
        
To sum up, \xnot.\xreq\ is the same as the hypothetical \xexcl\ when we know that the instance is an object.
Otherwise, the story is slightly more complex. 
\end{remark}

While \xreq\ usually occurs as a the only argument of \xnot, we found 16 cases 
(Table \ref{tab:notreq}, \complex\ cases) where it co-occurs with other
keywords, most typically with \xprops\ as in the following examples.

{\small     
\begin{verbatim}
"not": {"required": ["bundleDependencies"], 
        "properties": {"bundleDependencies": {}}}
          
"not": {"required": ["objectType"],
        "properties": {"objectType": {"enum": ["SubStatement"]}}}

\end{verbatim}
}

\code{
select sibkeys, count(*)
from treewithsiblings
where path like '
and sibnum > 1
group by sibkeys
order by count desc
}

The first example, characterized by the use of the trivial type \{\}, deserves closer inspection.

The semantics of \qprops: \{"\key{k}": \{\}\} is:
\begin{quote}
if the instance is an object then: if the instance has the \key{k} property then: the value of \key{k}
satisfies \{\}
\end{quote}
Since every instance satisfies \{\}, an assertion \qprops: \{"\key{k}": \{\}\} is always true,
hence 

{\small     
\begin{verbatim}
"not": {"required": ["bundleDependencies"], 
        "properties": {"bundleDependencies": {}}} 
\end{verbatim}
}

is just a verbose way of writing

{\small     
\begin{verbatim}
"not": {"required": ["bundleDependencies"] }
\end{verbatim}
}

The second case is subtler. Consider the following schema,
where $S$ is not trivial.

\qnot: \{\qreq: [\qkey{k}],  \qprops: \{\qkey{k}: $S$ \}  \}   --(\emph{notReqProS})

We first observe that if the instance is \emph{not} an object then the body

 \{ \qreq\ \ldots \qprops\ \ldots \} 
 
is trivially satisfied,
hence the entire (\emph{notReqProS}) fails. Hence the semantics of (\emph{notReqProS}) is 
\qtype\ : \qobject\ AND \ldots. 

We then observe that if the analysed object does not contain \key{k}, then the body fails, hence the 
entire (\emph{notReqProS})  holds,
hence its semantics is something like:
\qtype\ : \qobject\ AND (if \key{k} is present then \ldots).

Finally, for an object that contains the \key{k} field, the only way to fail  \key{k}\ : $S$ is to have the
\key{k} field bound to an instance that does not satisfy $S$, hence the specification above is equivalent 
to:

\qtype: \qobject, \qprops: \{ \qkey{k}:  \{  \qnot: S \} \} 

To sum up, we have the following equivalences:

(1)\\
\qnot: \{ \qreq: [\qkey{k}],  \qprops: \{ \qkey{k}:  \{ \} \} \} \\
$\ \Iff\ $
\qnot: \{ \qreq: [\qkey{k}] \} 

and

(2)\\
\qnot: \{ \qreq: [\qkey{k}],  \qprops: \{ \qkey{k}:  S \} \} \\
$\ \Iff\ $
\qtype: \qobject, \qprops: \{ \qkey{k}:  \{  \qnot: S \} \} 

In case (1), the \xprops\ specification is useless. 
The jargon of case (2) seems, to our eyes, a slightly more complicated way 
of saying, at the same time, that \key{k} should not satisfy $S$ and the instance should really be an
object.

We suspect that the reason why the first jargon is relatively common is because schema designers find it strange to have a
\xreq\ assertion without a related \xprops\ assertion nearby. For the second jargon, we have no intuition.

Besides the 173 cases where \xreq\ is bound to a unary array, we have found 64 situations 
(Table \ref{tab:notreq}, case ``*/twofields'') where 
\xnot.\xreq\ is followed by an array of two keys, as in the following example.

{\footnotesize
\begin{verbatim}
"not": {"required": ["result", "error"],
        "description": "cannot have result and error at the same time"}
}
\end{verbatim}
}

This schema requires the instance to violate \qreq: ["\key{result}","\key{error}"],
hence the instance must be an object and must miss one of the two fields, or both of them.
In other words, this jargon specifies a mutual exclusion constraint: if \key{result} is presente,
then \key{error} is absent, and vice versa.

Observe that "not": \{"required": ["a1", \ldots, "an"]\} woult not imply mutual exclusion among the $n$ fields,
but the fact that at most $n$-1 of them must be present, which seems somehow less useful.
Indeed, while positive intances of \xreq\ are followed by arrays of any length 
(we found one example with 411 entries), all instances of \xnot.\xreq\ that we found are followed
by either one key (field exclusion), two keys (mutual exclusion), or three keys (max two keys out of three),
but three keys only appear in three instances.

\code{
---three keys just two times
select line, pkey, key, jsonb_array_length(value) as numFields,value
from dftree
where path like '
      and jsonb_array_length(value) > 2

-- positive instance with 411 elements in the array
select forcelen(value), line, key, value
from dftree
where key = 'required'
order by forcelen desc
}

While the typical use of the \xnot.\xreq\ combination is the most elementary one in the direct case
(simple, one field),
the opposite happens when we consider the \xnot.\xderef.\xreq\ path, that is, the situation where
the negation of a \xreq\ assertion is mediated by a reference.\GG{We should delete all spurious
\xnot.\xit.\xnot.\xderef\ cases, but we do not have big changes, cases go from 36 to 25}
In this case, we have the results illustrated in Table \ref{tab:noterefreq}: we found no instance where
the negated referred object was simple, but in all 36 different instances is was complex, with a number
of assertions ranging from 3 to 7. In all 36 cases, \xreq\ was accompanied by a \xprops\ assertion
and, in 35 cases out of 36, also by a \qtype: \qobject\ assertion.  
In the same way, the number of fields whose simultaneous presence is excluded, in this case,
is higher than one in 10 cases out of 36 (Table \ref{tab:noterefreq}, cases \emph{two/manyfields}), and in one case it arrived
up to 7. 

\begin{table}\label{tab:noterefreq}
\center
\begin{tabular}{| l | l |r|  r  |}
\hline
Simple/complex & excluded fields & occurrences & files \\
\hline
complex	&	*	&	36	&	15	\\
complex	&	onefield	&	26	&	7	\\
complex	&	manyfields	&	7	&	5	\\
complex	&	twofields	&	3	&	3	\\\hline

\hline
\end{tabular}
\caption{The structure of arguments of \xnot.\xderef\ that contain \xreq.}
\end{table}

The negation of a \xreq\ assertion with a list of seven fields is quite a strange requirement, when considered in isolation: 
it is satisfied whenever the instance presents any of the $2^7-1$ strict subsets of the seven field names, 
but is violated if all the seven are present. 
In practice, we have observed that this kind of negative specifications are not used in isolation,
but rather in conjunctive contexts with the following structure, where \qkey{RefOne} establish a context 
that helps understanding the real meaning of \qnot\ : \{ \qdref : \qkey{RefTwo} \}:
{\small
\begin{verbatim}
"allOf" : [ { "$ref" :  "RefOne"} , {"not" : { "$ref"  :  "RefTwo"} } ]
\end{verbatim}
}\GG{This is not clear, needs an example}

This analysis confirms what we had already noticed from Table \ref{tab:notargs}: in our analysis we cannot just substitute
\qdref : \qkey{x} with the schema that is referred by \qkey{x}.
This substitution is correct from a semantic point of view, but is misleading in the context of usage
analysis. The data that we collected clearly indicate that the shape of the argument of negation is very different
when the argument is explicit and when the argument is mediated by a \xdref: in the first case the structure of the
argument is usually elementary, in the second case it is typically much more complex, in the first case what is negated is 
typically an assertion, in the second case is most commonly the definition of a class of objects.
In a sense, it seems that definitions are not used like macros that may be used in order to factorize any commonly
used combination of assertions, but rather to give a name to a set of homogeneous objects,
 like classes in an object-oriented language.
For this reason, we will always distinguish here between the immediate arguments of negation and those that 
are mediated by \xdref.

\subsubsection{\xnot.\xenum\ and \xnot.\xconst: value exclusion}\label{sec:notenum}

The path \xnot.\xenum\ appears 61 times and the path  \xnot.\xconst\ appears 6 times.
In the \xenum+\xconst\ case, the negated schema is usually \simple\ 
(44+6 cases out of 61+6, see Table \ref{tab:notenum}).
In the remaining 17 \complex\ cases, \xenum\ is always paired with a \qtype\ : \qstr\ assertion, which is
redundant since all values listed by \xenum\ in these cases are strings. This co-occurrence has little to do with 
negation, since also in the positive schemas \xenum\ is paired with a redundant \xtype\ assertion in the vast majority
of cases.

The list of excluded values is typically quite short, and 27 out of 61 negated occurrences of \xenum\ only exclude one value.

\begin{table}\label{tab:notenum}
\center
\begin{tabular}{| l | l | r|  r  |}
\hline
\multicolumn{2}{| l |}{\xnot.\xenum} & occurrences & files \\
\hline
*	&	*	&	61	&	52	\\
simple	&	*	&	44	&	36	\\
complex	&	*	&	17	&	17	\\
*	&	manyValues	&	29	&	24	\\
*	&	oneValue	&	27	&	25	\\
*	&	twoValues	&	5	&	5	\\
simple	&	manyValues	&	25	&	20	\\
simple	&	oneValue	&	14	&	12	\\
complex	&	oneValue	&	13	&	13	\\
simple	&	twoValues	&	5	&	5	\\
complex	&	manyValues	&	4	&	4	\\

\hline
\multicolumn{2}{| l |}{\xnot.\xconst} & occurrences & files\\
\hline
complex	&	oneValue	&	3	&	1	\\
simple	&	oneValue	&	3	&	1	\\\hline
\end{tabular}
\caption{Occurrences of \xnot.\xenum\ and \xnot.\xconst}
\end{table}

Hence, the following are the three typical cases \refschema{89480}, \refschema{5887}, \refschema{3458}.

{\small
\begin{verbatim}
"not": { "enum" : [ "markdown","code","raw" ] }

"not": { "const": "DateTime" }

"not": { "enum": ["generic-linux"], "type": "string" }
\end{verbatim}
}

Such schemas have an obvious interpretation: the instance may have any type and must be different from the string or strings that
are listed.

By expanding references, we get 12 more cases, where, as in the previous section, complex schemas
are more common than simple ones. In this case the numbers are too small to deserve a table.

\gcomment{
\begin{table}\label{tab:noterefenum}
\center
\begin{tabular}{| l | l | r|  r  |}
\hline
enum &&&\\
\hline
complex	&	oneValue	&	7	&	4	\\
complex	&	manyValues	&	3	&	2	\\
simple	&	oneValue	&	1	&	1	\\
simple	&	manyValues	&	1	&	1 \\
\hline
\end{tabular}
\caption{Occurrences of \xnot.\$xderef.\xenum}
\end{table}
}
\code{
--see above - uncomment 
	--path like '
-- and comment
   path like '
}

\subsubsection{\xnot.\xpatt\ }

The path \xnot.\xpatt\ appears 47 times, in 28 files. The \xpatt\ schema is simple in 47 cases over 48, which is surprising since,  
in the positive cases, \xpatt\ is almost invariably coupled with \qtype:\qstr\ (in approximately 17,900 cases out
of 18,300).
Under negation, however, schema designers prefer the structure that we exemplify below, where \xtype\  co-occurs 
with \xnot\ rather than with \xpatt. Observe that, differently from the \xenum\ case, here \qtype:\qstr\ is not redundant,
since \xpatt\ does not imply that the instance is a string, but is, on the contrary, satisfied by any instance that 
is not a string (and the same is true for \xformat). Hence, the choice of the presence, and the position, 
of \qtype:\qstr\ is not stylistic but semantic. The following is a typical example of negated pattern: the 
\xpatt\ schema is simple, but the \xnot\ is in a context that forces the type to \xstr, and that often presents 
other assertions \refschema{57662}.

{\small
\begin{verbatim}
{ "format": "uri",
  "not": { "pattern": "(\\$)" },
  "type": "string"
}
\end{verbatim}
}

Apart from the slightly different use of \xtype, the typical use of \xnot.\xpatt\ is very similar to the typical use of \xnot.\xenum: a
general string type is defined, and some exceptions are noted. In the following specific example, the \xnot.\xpatt\ assertion 
is indeed equivalent to a \xnot.\xenum\ assertion (thanks to the presence of the outer \qtype :\qstr) \refschema{51181}.

{\small     
\begin{verbatim}
{"type": "string",
 "description": "The definition_type property identifies ...",
 "not": { "pattern": "^(statement)|(tlp)$" }
}
\end{verbatim}
}

\code{
--pattern: counting the categories
select coalesce(issimple,'*'), '&', 
        count(*), '&', 
		count(distinct t.line) as numfiles, '\\',
	    array_agg(distinct sibnum) as lengths,
       jsonb_agg(distinct t.sibkeys) as keys,
        jsonb_agg(distinct t.value) as args
from treewithsiblings t,
     ift (sibnum = 1, 'simple'::text, 'complex'::text) as issimple
where path like '
group by cube(issimple)
order by case  when issimple is null then 3
			   else 4 end,
      count desc;
      
/*
*	&	47	&	28	\\
simple	&	46	&	27	\\
complex	&	1	&	1	\\
*/

--counting cooccurrences in positive cases (requires table ``cooccurrences'')
SELECT * 
FROM cooccurrences
where key1 in ('pattern')
order by count desc;

--checking how many times is paired with type: string
SELECT  t1.key, count(distinct t2.dewey), t2.key, array_agg(t2.value) 
FROM flattree t1 join flattree t2 on (t1.pdewey=t2.pdewey)
WHERE --t1.pkey='not' and 
      t1.key='pattern'  and t2.dewey != t1.dewey
group by  t1.key, t2.key
order by count desc

SELECT  count(*)
from dftree
where key = 'pattern'

--checking it never appears below $eref
select *
from edftree
where path like '

}

The path \xnot.\xderef.\xpatt\ --- that is, a pattern \xnot.\xdref\ where the reference denotes 
a type with a \xpatt\ member --- is not present in our collection.

\subsubsection{\xnot.\xtype: lot of redundant use}

The assertions that we have seen up to now are used to exclude some specific keys from an object type or some specific 
values from a string type. The use of \xnot.\xtype\ is somehow different, as we will see now.

When \xtype\ appears in a negated schema,we have three cases that appear with similar frequence:
negation of a \simple\ schema (21 occurrences),
of a \complex\ schema declaring a \xstr\ type (22 occurrences), of a \complex\ schema declaring an \xobject\ type
(19 occurrences).

\begin{table}\label{tab:nottype}
\center
\begin{tabular}{| l | l | r|  r  |}
\hline
IsComplex &	HowManyTypes & Occurrences & Files  \\
\hline
complex	&	one: \qstr & 22	&	20	\\
complex	&	one: \qobject & 19	&	17	\\
simple	&	one type	&	12	&	9	\\
simple	&	many types	&	9	&	7	\\
\hline
\end{tabular}
\caption{Occurrences of \xnot.\xtype}
\end{table}

\paragraph{simple\ schema}

When the negated schema is \simple, we have both single-type instances as the following one:

{\small
\begin{verbatim}
"not": { "type": ["number" ] }
\end{verbatim}
}
and many-types instances, as in \refschema{12174}:

{\small
\begin{verbatim}
"not": { "type": ["array","object","null" ] }
\end{verbatim}
}

At a careful analysis, negated \simple\ schemas of the single-type family are used in a redundant way in a surprisingly
high number of schemas.
Consider for example the following schema \refschema{35140}. 

(anyOfNotType)
{\small     
\begin{verbatim}
"anyOf" : [ { "not" : { "type": "array" } },
            { "items" : { "pattern" : .....} }
]
\end{verbatim}
}

By classical logical equivalences, it specifies that, if the instance is an array, then it must satisfy:

(oneLine)
{\small     
\begin{verbatim}
"items ": { "pattern" : .....} 
\end{verbatim}
}

However, the implication \emph{if the instance is an array} is already part of the semantics of \xit,
hence all the \xany-\xnot-\xtype\ part is redundant, and the schema (oneLine) is equivalent to the longer 
schema (anyOfNotType).

Another redundant use is the following one \refschema{2142}, 
where the final \xnot-\xtype-\xobject\ line could just be removed
since it is implied by the rest of the schema.

{\small     
\begin{verbatim}
{ "anyOf": [ { "type": "string",
      	     		  "pattern": "^Annotation$"
              },
              { "type": "array",
                "anyOf": [ { "type": "string",
                             "pattern": "^Annotation$"
                           } ]
              }
  ],
  "not": { "type": "object" }
}
\end{verbatim}
}

\hide{
Finally, consider the following example (15041, similar to f14983, 10999, f17145)
{\small     
\begin{verbatim}
"if": { "not": { "type": "object" } } ,
"then": { "type": "string", "enum": ["basic"]},
"else": { "properties" : ... }
\end{verbatim}
}
It could be easily rewritten as follows, with no use of negation

{\small     
\begin{verbatim}
"anyOf": [ { "type": "string", "enum": ["basic"]},
           { "type": "object", "properties" : ...   }
]
\end{verbatim}
}
}

We also found cases where the pattern is not redundant since it is used just to specify: ``this field cannot be a number'',
but the redundant cases seem as common as the non-redundant ones, if not more.

The \simple-schema many-types cases are rare but quite interesting. Consider, for example, the following schema \refschema{4179}. 
{\small
\begin{verbatim}
"not":{"type":["string","number","array","object","boolean","null"]}
\end{verbatim}
}
It may be surprising, since the authors are excluding all {\JS} types. Actually, they are requiring the use of a non-standard type.

\paragraph{\Complex\ schemas with \qtype:\qstr}

Out of 22 \complex\ schemas that declare \qtype:\qstr, 21 combine that declaration
with either \xenum\ or \xconst, hence making it redundant, as in the following example. 

{\small     
\begin{verbatim}
"not": { "enum": ["generic-linux"],
         "type": "string"
 }
\end{verbatim}
}

\code{
-- combining xtype-object with object assertions: 22 cases
SELECT  t1.dewey, t1.pkey, s.sibkeys as uncles, t1.key, t1.value, 
      array_agg(t2.key) as siblings, count(*) as sibnum
FROM flattree t1 join flattree t2 on (t1.pdewey=t2.pdewey)
     join treewithsiblings s on (t1.pdewey=s.dewey)
WHERE t1.pkey='not' and t1.key='type'  
      and t2.dewey != t1.dewey  --check that it is complex
	  and t1.value = '"object"'
group by  t1.dewey, t1.pkey, s.sibkeys, t1.key, t1.value --, t2.key
order by sibkeys

-- combining xtype-string with string assertions: 19 cases
SELECT  t1.dewey, t1.pkey, s.sibkeys as uncles, t1.key, t1.value, 
      array_agg(t2.key) as siblings, count(*) as sibnum
FROM flattree t1 join flattree t2 on (t1.pdewey=t2.pdewey)
     join treewithsiblings s on (t1.pdewey=s.dewey)
WHERE t1.pkey='not' and t1.key='type'  
      and t2.dewey != t1.dewey  --check that it is complex
	  and t1.value = '"string"'
group by  t1.dewey, t1.pkey, s.sibkeys, t1.key, t1.value --, t2.key
order by sibkeys

}

This redundancy is not related to negation: in the set of {\JS}s that we analyzed, half of the positive 
\xenum\ assertions 
are coupled with a redundant \qtype:\qstr\  assertion. 
Hence, the fact that 17 out of the 61 \xnot.\xenum\ paths are coupled with a 
redundant \xtype\ is just an instance of this general phenomenon, that we already discussed in Section \ref{sec:notenum}.

Only one schema \refschema{65826}. 
combines \qtype:\qstr\ with \xpatt. Due to the implicative semantics of 
\xpatt, adding a \qtype:\qstr\ to \xpatt\ is not redundant, and it implies that everything
that is not a string satisfies the negation. However, the assertion \qtype:\qstr\  that we 
find in the same schema at the outer level (see the schema below) forces the instance to be a string, hence, also in this case, the 
\qtype:\qstr\  assertion that is under negation (line 5) is redundant.

{\footnotesize     
\begin{verbatim}
"items": {
          "type": "string",
          "minLength": 1,
          "not": {
                   "type": "string",
                   "description": "...string that contain two or more (*)",
                   "pattern": ".*\\*.*\\*.*"
          } ...
\end{verbatim}
}

\paragraph{\Complex\ schemas with \qtype:\qobject}

We now analyse the 19 negated \complex\ schemas that combine \qtype:\qobject\ with object assertions.
While \qtype:\qstr\  is redundant when combined with an \xenum\ assertion that only enumerates
strings, \qtype:\qobject\ would \emph{not} be redundant, in general, near to an object assertion,
due to the hypothetical semantics of {\JS}. While the following assertion

\qreq : [ $k$ ]

means: \emph{if} the instance is an object, \emph{then} $k$ is required, the combination:

 \qtype : \qobject, \qreq : [ $k$ ]
 
means: the instance \emph{is} an object \emph{and} $k$ is required, and this distinction is reversed under negation, as discussed in Remark \ref{rem:implication}. 

However, like in the above \xpatt\ example, the \qtype:\qobject\ part becomes redundant if it is already asserted 
in the positive part of the schema, as in the following example \refschema{24167}.

{\small     
\begin{verbatim}
{ "not": { "type": "object",
           "properties": { "scheme": { ... } }
         },
  "type": "object"
}
\end{verbatim}
}

The structure above is found in 13 out of 19 negated \complex\ schemas that contain \qtype:\qobject.

\code{
-- see the ``uncles'' field of the code above
}

\paragraph{\xnot.\xderef.\xtype} 

The situation is totally different with the path \xnot.\xderef.\xtype, that is, when we have  \xnot.\xdref\ that 
refers to a schema that contains \xtype. We have 51 instances, all of them \complex\ and all of them with
just one type, and only in 5 cases out of 51 the negation co-occurs with a \xtype\ as in the example above.\GG{21 of these 51 cases are at the end of \xnot.\xit.\xnot, they are not real negations}
In 10 case out of 51, the path \xnot.\xderef\ arrives at a combination \qtype:\qstr\ plus \xenum, which makes
\qtype:\qstr\ redundant, but in the other 41 cases \xnot.\xderef\  arrives at a combination 
\qtype:\qobject\ plus some object keywords, usually \xprops\ and \xreq, so that finally  
\qtype:\qobject\  is not redundant: thanks to its presence, an instance that is not an object satisfies this
specification.\GG{The same happens if you }

To sum up, under negation, \xtype\ is used either as \qtype:\qobject\ in a \complex\ object schema, or as
\qtype:\qstr\ in combination with \xenum, or as a \simple\ schema by itself.
While in the \xenum\ case it is always redundant, in the other two cases we have found a mixture of redundant and
non-redundant use - typically non redundant when mediated by \xderef\ and when \xtype\ is followed by an array,
typically redundant in the other cases.
Of course, redundancy is not necessarily a problem - redundant assertions may be added on purpose for reasons
of homogeneity or for readability.


\subsection{Negation of applicators: boolean operators, object operators, array operators}\label{sec:applicators}

\subsubsection{Negation  of boolean operators: general introduction}

The three boolean combinations \xnot.\xany, \xnot.\xall\ and \xnot.\xone\ all appear with a non-negligible
frequency. The first thing we measure is the relative frequency, and it is interesting to observe that here \xone\ is not any more the most-used combinator,
but becomes the less used one (Table \ref{tab:notanykey}).
This makes sense: when the aim is reasoning on the negation of boolean operators,
\xnot.\xany\ is easier to understand than \xnot.\xone. \custcom{\tiny also, there are probably more use natural cases for not.anyOf than not.oneOf}

While most negated schemas are simple, we have 55 of them that are complex. 
A \complex\ schema \{ \qkey{a} : S, \qkey{b} : T \} is equivalent to a conjunction of \simple\ schemas
\qall [  \{ \qkey{a} : S \} , \{ \qkey{b} : T \} ], hence we will analyse complex schemas together with
boolean conjunctions, although there is an important difference: in a complex schema, all fields hace a different
keyword, while in a \xall\ the same keyword may appear in many different arguments. As we will see, this 
difference is very important in practice.

\code{
--distinguish instances of not on the basis of arg size
select
(select count(*) from treesibargs
	where key = 'not' and valuelennew > 1),
(	select count(*) from treesibargs
	where key = 'not' and valuelennew = 1),
(	select count(*) from treesibargs
	where key = 'not' and valuelennew = 0),
(	select count(*) from treesibargs
	where key = 'not' )
	from dual
	
--count not.$ref
select key, count(distinct pdewey), count(*)
from edftree
where pkey = 'not' and key like ('$_ref')
group by cube(key)

--count complex negated references: 79 su 93
select count(*)
from
(select jsonb_agg(e.key), count(*)
from eflattree e 
where e.p2key = 'not' and e.pkey = '$eref'
group by pdewey
having count(*) > 1)

}

In presence of a boolean combinator, the next natural question is which keyword do we typically find below.
In Table \ref{tab:notanykey} we count the instances of \xnot.(\xany/\xone/\xall)[*].$k$ for the different keywords.

\begin{table}\label{tab:notanykey}
\center
\begin{tabular}{| l | l | l | r|r|  r  |}
\hline
        &op &  $k$ & {\xdss}not.op\fb\xarrs.$k$ & {\xdss}not.op\xarrs.$k$ & Files \\
\hline
not	&	anyOf	&	required	&	27	&	71	&	22	\\
not	&	anyOf	&	\$fref	&	8	&	18	&	7	\\
not	&	anyOf	&	\$eref	&	5	&	13	&	4	\\
not	&	anyOf	&	properties	&	4	&	10	&	3	\\
not	&	anyOf	&	pattern	&	1	&	2	&	1	\\
not	&	anyOf	&	description	&	1	&	2	&	1	\\
\hline
not	&	allOf	&	required	&	23	&	46	&	23	\\
\hline
not	&	oneOf	&	\$eref	&	4	&	6	&	2	\\
not	&	oneOf	&	properties	&	2	&	4	&	2	\\
not	&	oneOf	&	required	&	2	&	2	&	2	\\
not	&	oneOf	&	\$fref	&	2	&	2	&	1	\\
\hline
\end{tabular}
\caption{Number of occurrences of \xnot.\xany/\xone/\xall[*].$k$ for the different keywords $k$}
\end{table}

\code{
-- Number of occurrences of \xnot.\xany/\xone/\xall[*].$k$ for the different keywords
Select e1.p3key, '&', --p2s.sibkeys as uncles, 
        e1.p2key,  '&', e1.key, '&', 
		count(distinct e1.p2dewey) as "not.op([where *])", '&',
		count(*),  '&',
		count(distinct e1.line) as lines,
		 '\\'
FROM eflattree e1 
WHERE e1.p3key='not' --and e1.pkey='anyOf' 
      and not e1.added
      AND e1.pkey similar to '[0-9]+'
--and sibnum = 1
GROUP BY e1.p3key, e1.p2key, e1.key
ORDER BY e1.p2key, count desc, lines desc
}

As expected, \xreq\ is still the dominating case. This will be discussed below.


\subsubsection{Negation of \xany}\label{sec:notany}

\code{
--notanyOf how many: 45
select count(*)
from dftree
where path like '

-----the following query visualized all not.boolOp collecting the set of keywords below
select  op.o, a.line, a.num as argnum, a.arg, 
      count(*) as children, 
	  count(distinct (op.o, a.line, a.num)) as notArgsNum,
	  count(distinct a.line) as lines, 
      array_agg(distinct k.keyword),
	  array_agg(distinct arg.opArg)
from notargs a , keywords k, 
     ( values ('anyOf'),('oneOf'),('allOf')) op (o),
	 jsonb_path_query(a.arg, concat('$.', op.o, '[*].',k.keyword)::jsonpath)
	 with ordinality as arg (opArg,num)
where a.arg @? concat('$.',  op.o, '[*].',k.keyword)::jsonpath 
and k.kclass not in ('array', 'path')
and o = 'anyOf'                                    --------modify this
and k.keyword = 'required'
group by op.o, a.line, a.num, a.arg      ----modify this
having 'required' =any (array_agg(distinct k.keyword))    --------modify this
order by count(*) desc,  op.o, array_agg(distinct k.keyword)
}

\code{
--This query shows that, out of 50 occurrences of not.any, each of them is either
--followed by ref or by properties or by required
select line, a.anum, array_agg(distinct k.key), count(*)
from dfn2,
     jsonb_path_query(sch,'strict $.**.not.anyOf')
	 with ordinality as a (a,anum),
	 jsonb_array_elements(a.a)
	 with ordinality as e (elem, enum),
	 jsonb_object_keys(e.elem) 
	 with ordinality as  k (key, keynum)
group by line, anum;

}

The most striking feature of the occurrences of \xnot.\xany\ is the extreme homogeneity and simplicity
of the arguments of the disjunction. We found 46 instances of this pattern, and in each of them all
the arguments of the disjunction have exactly the same set of operators, and that set is always a singleton,
apart from one case where the arguments present two keys (\xdescr\ and \xreq). 
This is ilustrated in Table \ref{tab:notany}: in 26 cases, every argument of \xnot.\xany\ 
is a simple schema that only contain a \xreq\ keyword.
In 13 cases, every  argument is a simple schema that only contain a \xdref\ keyword.
In 4 cases, every  argument is a simple schema that only contain a \xprops\ keyword.
In 1 case, each of the two arguments contains both \xdescr\ and \xreq.
In 1 case, each of the two arguments contains \xpatt.

For each keyword $k$, the first column in the table reports the occurrences of \xnot.\xany\ that are followed
by $k$, and the second column the number of occurrences of \xnot.\xany\xarrs.$k$. 
We see for example that 26 occurrences of \xnot.\xany\ followed by \xarrs.\xreq\ correspond to 69 occurrences of 
\xnot.\xany\xarrs.\xreq, which means that, on average, every \xany\ has 69/26 arguments, which means that, in 
practice, most disjunctions have just 2 arguments.

\begin{table}\label{tab:notany}
\center
\begin{tabular}{| l | r| r| r| r  |}
\hline
$k$  &   \xdss\xnot.\xany\fb\xarrs.$k$ &  \xdss\xnot.\xany\xarrs.$k$  &  files  \\
\hline
{required}	&	26	&	69	&	26	\\
{\$ref}	&	13	&	31	&	13	\\
{properties}	&	4	&	10	&	4	\\
{description,required}	&	1	&	2	&	1	\\
{pattern}	&	1	&	2	&	1	\\
\hline
\end{tabular}
\caption{number of occurrences of \xnot.\xany\ and of \xnot.\xany.$k$, grouped according to the set of keys that is reached 
starting from the target of \xnot.\xany}
\end{table}

\code{
-- computing the table
with temp as --temp: one line for each not.anyOf
(select --p2dewey, p3key, p2key, array_agg(pkey) pkeys, 
        array_agg(distinct key) keys,
        count(distinct pdewey) as pcount,
        count(distinct dewey) as numarg,
        count(distinct line) as lines,
        array_agg(distinct line) as linelist
from flattree --eflattree
where p3key = 'not' and p2key = 'anyOf' --and added = false
group by p2dewey) --, p3key, p2key)
select keys, '&', count(*), '&', sum(pcount), '&', --sum(numarg), '&', 
      sum(lines), '\\', array_agg(distinct linelist)
from temp
group by rollup(keys)
order by count desc;
}

Hence, the  \xnot.\xany\xarrs.\xreq\ path constitutes the vast majority of the uses of \xnot.\xany, and they
are typically very simple, as in the following example, where \xany\ has two arguments and each 
\xreq\ has one.

{\small     
\begin{verbatim}
"not": {
    "anyOf": [
        { "required": ["constructor"] },
        { "required": ["statics"] }
    ]
}
\end{verbatim}
}

This case $2\times 1$ case is the most common one. The number of branches of \xany\ varies actually from 1 to 10.
As for the number of arguments of \xreq, it is either 1 when we want to express
field exclusion or 2 when we want to express mutual exclusion, as in Section \ref{sec:notreq}. 
In both cases, the number of arguments of \xreq\ is always homogeneous at the internal of the same \xany, and  we
have 17 cases with 1 argument in every branch (field exclusion) and 10 cases with two arguments (mutual exclusion) in every
branch.

\code{
---look all the lists
with temp as
(select count(*) as numRequiredArgsOfAnyOf,
        array_agg(value) as skeleton,
        jsonb_agg(len(value)) as listOfRequiredListLength,
        min(len(value)) as minRequiredListLength,
        max(len(value)) as maxRequiredListLength
from flattree
where p3key = 'not' and p2key = 'anyOf' and key = 'required'
group by p2dewey)
select listOfRequiredListLength, minRequiredListLength,
       len(listOfRequiredListLength), count(*)
from temp
group by listOfRequiredListLength, minRequiredListLength
order by count(*) desc;

--generate numbers: when reqlist = 1, we have 10 cases; when reqlist = 2, we have 17 cases; 
with temp as
(select count(*) as numRequiredArgsOfAnyOf,
        array_agg(value) as skeleton,
        jsonb_agg(len(value)) as listOfRequiredListLength,
        min(len(value)) as minRequiredListLength,
        max(len(value)) as maxRequiredListLength
from flattree
where p3key = 'not' and p2key = 'anyOf' and key = 'required'
group by p2dewey)
select minRequiredListLength as reqListLen,count(*),
       max(maxRequiredListLength) as reqListLenControl,
       min(len(listOfRequiredListLength)) as minLenAnyOfArg,
	   max(len(listOfRequiredListLength)) AS mAXLenAnyOfArg
from temp
group by minRequiredListLength
order by count(*) desc;
}

In the first group, composed of 17 negated disjunctions of unary \xreq, the designer uses this combination to exclude from the instance all keywords in a set.
In the second group, of 10 negated disjunctions of binary \xreq, the \xnot.\xany\ construct is used to collect a set of 
$n$ mutual exclusion statements. 
By direct analysis, we have seen that, in many cases, these $n$ statements describe one field that is exclusive wrt to $n$ other 
fields, in other cases they collect $n$ pairs regarding 2*$n$ distinct fields, but there are rare cases that describe
exclusions graphs that are more complex than these.

The \xnot.\xany.\xderef\ case is similar to the previous one in the sense that, if one analyses the set of references
collected below the same \xany, they are always extremely homogeneous.
They are all \complex\ schemas that either combine \qtype:\qstr\ with 
\xpatt\ or \qtype:\qobject\ with \xprops\ and \xreq.

\code{
--going through $eref (5 cases with 13 references)
with temp as --temp: one line for each not.anyOf
(select --p2dewey, p3key, p2key, array_agg(pkey) pkeys, 
        --array_agg(distinct key) keys,
        --count(distinct pdewey) as pcount,
        --count(distinct dewey) as numarg,
        count(distinct line) as lines,
        jsonb_agg(value) as values,
        array_agg(distinct line) as linelist
from eflattree
where p3key = 'not' and p2key = 'anyOf' 
      and key = '$eref'
group by p2dewey) --, p3key, p2key)
select --keys, '&', count(*), '&', sum(pcount), '&', --sum(numarg), '&', 
      count(*),
	  forcelen(values),
	  (values),
	  sum(lines), '\\', array_agg(distinct linelist)
from temp
group by values
order by count desc;
}

The other cases, \xnot.\xany\ followed by \xprops\ and \xpatt\ are exceptional, and we will
not discuss them.

To sum up, \xnot.\xany\ is mostly used to combine fields exclusion statements, and is characterized by a short list of simple
and homogeneous arguments. 

\subsubsection{Negation of \xall}

The next most common case is \xnot.\xall\xarrs.\xreq. We found 23 occurrences.
All occurrences of \xnot.\xall\ only present \simple\ \xreq\ schemas as arguments.
They \emph{all} follow the same binary pattern below, the only variation being in the X and Y names
 \refschema{3212}, \refschema{9186}, \refschema{3020} \dots.

\code{-- count 23
select count(*)
from dftree
where path like '
}

{\small     
\begin{verbatim}
"not" : { "allOf" : [{"required": ["X"]}, {"required": ["Y"]}] }
\end{verbatim}
}
By and/or duality, we are here requesting to satisfy any of the two exclusion schemas, or both
(and, as usual, the instance must be an object).
That is, the only combination that is forbidden is when both $X$ and $Y$ are present, any other combination is
good. Hence the schema above is equivalent to the following schema (discussed in Section \ref{sec:notreq}).

{\small     
\begin{verbatim}
"not" : { "required": ["X", "Y"] }
\end{verbatim}
}

We feel that both specifications are not easy to read, but the second one is at least more compact, hence we wonder 
why the first one has been preferred in these 23 occurrences.

\code{
--verify the shape of allOf
select line, p3key, p2key, jsonb_agg(key), jsonb_agg(value)
from flattree
where p3key = 'not' and p2key = 'allOf' and key = 'required'
group by line, p3key, p2key, p2dewey

}

%

\subsubsection{Negation of \complex\ schemas}

A \complex\ schema \{ \qkey{a} : S, \qkey{b} : T \} is equivalent to a conjunction of \simple\ schemas
\qall [  \{ \qkey{a} : S \} , \{ \qkey{b} : T \} ], but is used in a very different way.

Negated complex schemas belong, essentially, to three different categories.
They either combine \xenum, \xconst, or \xpatt\ with \qtype: \qstr (22 occurrences).
Or, they combine a subset of \xprops, \xreq, \qtype:\qobject (21 occurrences).
Or, they combine \xaddProps, \xpattProps\ and, sometimes, also \qtype:\qobject (11 occurrences).
All these three combinations have been already discussed.

We only have one exception, where the negated combined schema also includes a \xany\ operator,
which changes completely the picture. This happens in schema \refschema{37789}
, and will be discussed in 
Section \ref{sec:implication}.

\begin{table}\label{tab:complex}
\center
\begin{tabular}{| l | r| r| r| r  |}
\hline
Structure &   occurrences  \\
\hline
\xenum, \xconst, \xpatt, \xtype:\xstr 	&	22	\\
\xprops, \xreq, \xtype:\xobject 	&	21	\\
\xaddProps, \xpattProps, \xtype:\xobject	&	11	\\
\xany, \xprops, \xreq 	&	1	\\
\hline
\end{tabular}
\caption{Combinations of keywords in }
\end{table}

\subsubsection{Negation of \xone}

The last boolean negation to consider is \xnot.\xone. 

This case is rare, since we only have six instances.
Moreover, in four cases the pattern \xnot.\xone\ is found at the end of a \xnot.\xit.\xnot.\xone\ pattern, 
and we believe, as discussed in Section \ref{sec:notitems}, that they must be understood as (\xnot.\xit.\xnot).\xone.
Hence, we only have two \emph{real} instances, but we will reserve them some space nevertheless, since
they give us an excellent example of how the interpretation of a JSON Schema may be complicated by the interplay
between negation, \xone, and implicative semantics (that is, the fact every number assertion is trivially satisfied 
by anything that is not a number, and the same for strings, arrays, and objects).
Both instances are binary, as in \qnot : \{ \qone : [ $S1, S2$ ] \} (as are the four \xnot.\xit.\xnot.\xone\ instances that we discarded).

\code{
--not.oneOf: 6 cases
select *
from eflattree
where path like '
}

We first observe that the semantics of \qnot : \{ \qone : [ $S1 \ldots Sn$ ] \} is: ``either all of $S1 \ldots Sn$ are not satisfied,
or exist $i,j$ with $i\neq j$ such that $Si$ and $Sj$ are both satisfied''.
This is quite complicated to understand, but in practice we noticed that it is often the case that all arguments [ $S1 \ldots Sn$ ] of any \xone\ operator, negated or not,
are mutually exclusive. In this case, \xone\ is equivalent to \xany, and \xnot.\xone\ is equivalent to \xnot.\xany,
which makes it much easier to understand.

However, deciding whether the different branches of \xone\ are exclusive is not obvious, since it often depends
on the context.

Consider the following example \refschema{4896}

{\small
\begin{verbatim}
"oneOf": [
    {   "properties": {
             "status": { "enum": ["content error" ] }
        },
        "required": [ "content_error_detail" ]
    },
    {   "not": {
            "oneOf": [
                { "properties":
                      { "status": { "enum": [ "uploading", "released" ]  }
                      },
                  "required": [ "content_error_detail"]
                },
                { "properties": 
                      {  "status": { "enum": [ "content error"] }
                      }
                }
    ]}}
]
\end{verbatim}
}

\gcomment{
 A XOr Not ( B XOr C) =  A XOr ( B Iff C) 
means: either the three are false, or exactly two are true.
0  XOr 00/11 =  0 XOr 1  => 1
1  XOr  01/10  = 1 XOr  0 => 1

0  XOr 01/10 =  0 XOr 0  => 0
1  XOr  00/11  = 1 XOr  1 => 0

Hence we have four cases, 000, 011, 101, 110. If we have status, then 11x is impossible and x11 is impossible,
hence we only have 000 and 101

Hence it means: not (S:UR and CED!), i.e. S:UR => not CED!
and C:C IFF C:C and CED!, that is:  S:C => CED!

If we have no other cases then U,R,C, then it means : ( CED! ? S:C else S:UR)
}

Let us analyse the internal \xnot.\xone.
Here, the two branches are \emph{not} mutually exclusive. They are both satisfied by anything that 
is not an object, and also by any object that has the required field and does not have the \emph{status}
field, hence this \xnot.\xone\ specification can be satisfied either by violating both constraints (as for
\xnot.\xany) but also by satisfying both constraints, which is quite bizarre.

But let us look now at the context around the \xnot.\xone.
The \xnot.\xone\ is combined here, through an external \xone, with a specification that is satisfied, because of
the implicative semantics of \xprops\ and of \xreq, by anything that is not an object,
and by any object that has the \emph{content\_error\_detail} and does \emph{not} have a \emph{status} field.
Hence, the exculsive semantics of the external \xone\ together with the first branch prevents exactly those cases 
that would make the internal \xnot.\xone\ different from \xnot.\xany.
For example, a number would satisfy both arguments of the internal \xone, hence it would violate the \xone\ itself,
hence it would satisfy the \xnot.\xone. But a number would also satisfy the first branch of the external \xone,
hence it would fail the entire specification. In the same way, an object with the \emph{content\_error\_detail}
field (hereafter, \emph{c\_e\_d}) and without a \emph{status} field would satisfy the three branches, hence would fail 
the entire specification.

So the internal \xnot.\xone\ is equivalent to \xnot.\xany\ because of the exclusive semantics of the external \xone. 
But if we zoom out a bit in the original schema (not reported here), we discover that the entire construct is inside a \xdeps : \emph{status} declaration,
and is combined with a \xall/\xone/\xany\ combination that forces the value of \emph{status} to be one of the 
three value listed (sctually they are much more than three, we are here simplifying the things a little bit),
so that the semantics of the \xone\ schema is relevant only for situations where the instance is actually an object that 
contains the \emph{status} field, with a value chosen from those listed. 
So, in order to decode this specification, we can restrict ourselved to this class of instances.

In practice, we find this reasoning technique extremely complex, because of the combination between implicative 
semantics and exclusive semantics. We have the impression that the simplest way
of decoding this specific schema is just a model-checking
approach, where we check which combinations of values for the \emph{status} field and presence/absence for the
\emph{c\_e\_d} field satisfy the specification. We propose the following alternative way of expressing the
entire specification, and leave to the reader the task to verify its equivalence with the original one, under the assumption 
that the instance is an object with a \emph{status} property and whose value is included in the list.

{\small
\begin{verbatim}
"oneOf": [
    {    "required": ["content_error_detail"],
         "properties": {
             "status": {"enum": ["content error" ]}
         }
    },
    {    "not": { "required": ["content_error_detail"]},
         "properties": {
             "status": { "enum": ["uploading", "released"]}
         }
    }
]
\end{verbatim}
}

The aim of this long analysis was to illustrate the 
complexity of the interplay between \xnot, \xone, and implicative semantics, which we believe explains why \xone, which
is the most common boolean operator in positive position, becomes the less frequent one when we move to the negated 
occurrences.

\subsubsection{\xnot.\xprops}

The path \xnot.\xprops\ appears 71 times. A negated schema that contains \xprops\ is \simple\ in 54 cases
over 71,
and the schema associated to \xprops\ contains only one property in 62 cases over 71, hence the typical case is the
\simple\ \xprops\ schema with one property only, as in the following example.

{\small
\begin{verbatim}
"not": { "properties": {"status": { "enum": ["released","revoked"]} } }
\end{verbatim}
}

The meanings of \xprops\ is implicative: if the instance is an object, if the field is present, then it must satify the
indicated subschema. Hence, its negation is conjunctive: the instance \emph{is} an object, the
property \key{status} \emph{is} present, and it must violate the subschema.
Hence, this simple specification combines the effect of \qtype : \qobject, \qreq: [\qkey{k}], and 
a negated type assertion for the property.
While this fact is important in theory, in practice we have seen this pattern mostly used 
in situations where the context already implies that the instance is an object with that field. 
In those situations, the position of the negation is not relevant, and we have the following equivalence:

{\small
\qtype: \qobject,
\qreq: [ \qkey{p} ], 
\qnot: \{ \qprops: \{\qkey{p}: $S$ \} \} $\iff$ \\
\qtype: \qobject,
\qreq: [ \qkey{p} ], 
\qprops: \{\qkey{p}: \{ \qnot: \{  $S$ \} \} \}
}

Hence, \xnot.\xprops\ is mostly used to impose that a member has a type \xnot\ $S$.
While negation is, in general, mostly used in front a \qreq, in the \xnot.\xprops\ case the type operator that is most often
negated is \xenum, as in the example above: \xnot.\xprops\ is mostly used to exclude 
some values from the domain of the property.
In the other few cases, the schema associated with $S$ 
features \xprops\ again, \xtype, some user-defined keywords, and also a \xmaxIt\ = 0, whose negation 
looks like a more complex way of expressing \xminIt\ = 1.

\code{
--keywords that cooccur with properties under negation
select line, pdewey, e1.value, jsonb_agg(e2.key), jsonb_agg(e2.value)
from edftree e1 join edftree e2 using (line,pdewey,added,pkey)
where pkey = 'not' and e1.key = 'properties'
       and e2.dewey != e1.dewey
	   --and e2.key = 'required' 
	   and not added
group by line, pdewey, e1.value --, e2.value
order by jsonb_agg(e2.key)
}

In the 17 \complex\ cases, \xprops\ is mostly combined with \xreq\ (11 cases over 17), as in the following example \refschema{76666}. 
{\small
\begin{verbatim}
"not": { "properties": { "objectType": {"enum": ["SubStatement"]} },
         "required": ["objectType"]
}
\end{verbatim}
} 
In the other cases, it is combined with \qtype:\qobject, or with both \xreq\ and \qtype:\qobject.

Each of these forms has a different meaning, for the reasons that we discussed many times, since we have the two following
equivalences, that specify that whenever one side forces the instance to be an object (is \emph{assertive} on that) its
negation is satisfied by any instance that is not an object (is \emph{implicative} on that feature), and the same with
\xreq: if a side is assertive on the field presence, the negation is implicative.

\xnot : \{ \qtype: \qobject,
\qreq: [ \qkey{p} ], 
 \qprops: \{\qkey{p}: $S$ \}  \\
\text{ } $\ \qquad\iff$ \{ \qprops: \{ \qnot: \{\qkey{p}: $S$ \} \} \\[2ex]
\xnot : \{ 
\qreq: [ \qkey{p} ], 
 \qprops: \{\qkey{p}: $S$ \}   \\
 \text{ } $\ \qquad\iff$ \{ \qtype: \qobject, \qprops: \{ \qnot: \{\qkey{p}: $S$ \} \} \\

Hence, the fact that we find all different combinations may give the impression that JSON Schema users have a sophisticated
usage of the assertive/hypothetical distinction. However, from our direct analysis we got rather the opposite impression: in
most situations, some contextual information forces the type to be an object, and sometimes forces the property to be
present, hence making all the variations above mutually equivalent.

\gcomment{
WHAT FOLLOWS IS OUTDATED BY THE LATEST OBSERVATIONS
The field that , \xprops\ is typically
associated with the trivial type \{ \}, as in the following example (f15041).

{\small
\begin{verbatim}
"not": { "properties": { "rejectVlans": {} },
            "required": ["rejectVlans"]
}
\end{verbatim}
} 

In all these situations, the line \xprops : \{ \key{k} : \{ \} \} is just redundant, since any JSON value satisfies that
specification, hence the above schema is equivalente to the following one.

{\small
\begin{verbatim}
"not": { "required": ["rejectVlans"] }
\end{verbatim}
} 

This habit of joining a useless \xprops\ to the relevant \xreq\ has already been discussed in Section \ref{}. 

Much more interesting are the schemas that combine \xreq\ with a non-trivial type under \xprops, as the following one (f23477, nicolaasmatthijs).

{\small
\begin{verbatim}
"not": {
                    "required": [
                        "objectType"
                    ],
                    "properties": {
                        "objectType": {
                            "enum": [
                                "SubStatement"
                            ]
                        }
                    }
                }
\end{verbatim}
}

As discussed in Remark \ref{rem:implication}, this combination means: the instance is an object and, \emph{if}
\key{objectType} is present, then its type must violate the type specified - in this case, its value must 
differ from "SubStatement". This is a quite sophisticated use of negation.
}

We conclude with yet another example that shows how negated schemas can be tricky to interpret, even when short; we leave this
one to the reader with no comment\GG{Actually, this one is quite simple to interpret}
\refschema{21864}.
\footnote{The schema is satisfied by any instance that is an object and either misses one of
the two members, or it has both members but at least one is not of type string.
}

 {\small
\begin{verbatim}
"not": { "properties": {
             "FirstName": {"type": "string"},
             "LastName": {"type": "string"}
         },
         "required": ["FirstName","LastName"]
}
\end{verbatim}
}

\subsubsection{\xnot.\xpattProps\ }

The path \xnot.\xpattProps\ is present in 15 schemas, and it follows two different usage patterns.
In 9 schemas, most of which seem to have been produced by the same group, we find the following structure \refschema{11675}.

{\small
\begin{verbatim}
"type": "object",
"patternProperties": {
     "^([0-9]{3})$|^(default)$": {"$ref": "#/definitions/responseValue"},
     "^x-": { "$ref": "#/definitions/vendorExtension" }
 },
 "not": {"type": "object",
         "additionalProperties": false,
         "patternProperties": {
                    "^x-": {"$ref": "#/definitions/vendorExtension" }
         }
"additionalProperties": false,
"minProperties": 1
 \end{verbatim}
}

\code{
--counting
select count(*), count(distinct line)
from dftree
where path like '

--having a look at the 15 different cases
select pkey, key, value, sibkeys, count(*)
from treewithsiblings
where path like '
group by pkey, key, value, sibkeys

--analisyng the contexts
--grouping pproperties on the basis of the pattern
--keywords that cooccur with properties under negation
-- e2: sibling of patternProperties; e2.dewey != e1.dewey to count the complex cases
--notsib: siblings of the not
select count(*), array_agg(e1.line) as files, --e1.pdewey, 
       forcelen(e1.value) as numProp,
       e1.value as ppArg, jsonb_agg(e2.key order by e2.key) as cooccur, 
      jsonb_agg(e2.value order by e2.value) as cooccurArg,
	  jsonb_agg(notsib.pkey) as notsibpkey,
	  jsonb_agg(notsib.key) as notsibkey,
	  notsib.value
from flattree e1 join dftree e2 using (line,pdewey,pkey) --added,
     join dftree notsib on (notsib.pdewey=e1.p2dewey)
where e1.pkey = 'not' and e1.key = 'patternProperties'
       and e2.dewey != e1.dewey
	   and notsib.key = 'patternProperties'
	   --and e2.key = 'required' 
	   --and not added
group by --e1.line, e1.pdewey, 
         e1.value, notsib.value --, e2.value
order by e1.value --jsonb_agg(e2.key)
}

This structure is interesting to analyse, to see how programmers use {\jsonsch}.
We first recall that, in {\jsonsch}, \xaddProps\ receives its meaning by the \xprops\ and \xpattProps\ that
co-occurr as members of the same object, since it specifies a schema for everything that does not match the cooccurring
\xprops\ and \xpattProps\ specifications.
The \qtype:\qobject\ inside the \xnot\ is redundant, because of the external \qtype:\qobject,
as we have discussed many times. The reference \emph{\#/definitions/vendorExtension} refers
to a trivial schema that is satisfied by any instance, hence the \xpattProps\ under negation is not used to limit the 
structure of the members whose name starts with \emph{x-}. Hence,  \xpattProps\ is just used to specify that 
the co-occurring \xaddProps\ refers to any member that does not start with \emph{x-}.
The negated schema describes an object where every field starts with \emph{x-}.
Hence, since the external schema forces
the instance to be an object, the only way of failing this negated schema is by presenting a member that does not start with \emph{x-}, and is associated to any value.
If we combine this with the pair \xpattProps-\xaddProps\ at the outer level that specifies that every 
field must match either the 
\kw{\NN[0\text{-}9]\{3\}\$|{\NN}default\$} (\emph{responseValue}) or the \kw{\NN\text{-}x} (\emph{vendorExtension}) specification, then this is a way to require that at least one member matches \emph{responseValue}.
By the way, observe that this makes the \xminP = 1 specification redundant. 

We find this structure extremely interesting: {\jsonsch} has a \xreq\ specification that allows the programmer to require the presence of a member with a given name, but has no \xpattReq\ operator to require the presence of member whose name matches a
specific pattern, hence the need of the complex construction above.
\gcomment{
We just observe that a slightly more direct way of asking for the presence of a field that matches that pattern would have been
the following one: \xpattProps\ $p$ : \xfalse\ is satisfied by all and only the instance that do not have a member that 
matches $p$, hence its negation imposes the presence of such a member.

{\small
\begin{verbatim}
"type": "object",
"patternProperties": {
     "^([0-9]{3})$|^(default)$": {"$ref": "#/definitions/responseValue"},
     "^x-": { "$ref": "#/definitions/vendorExtension" }
 },
 "not": { "patternProperties": {
                        "^([0-9]{3})$|^(default)$": false
            }
"additionalProperties": false,
 \end{verbatim}
}
}

}

\code{
}

Other 2 schemas, which seems to be different versions of the same schema, present instead the following structure
\refschema{64748} and \refschema{51189}.

{\small
\begin{verbatim}
"type": "object",
"description": "Specifies any other header fields (except for date, 
                   received_lines, .., and subject) found in the...",
"not": { "additionalProperties": false,
         "patternProperties": 
           { "^date|received_lines|subject$": 
                { "description": "Invalid additional header field types"
                }
            }
},
...
\end{verbatim}
}               

As before, since the schema associated to the internal pattern is trivially satisfied, the argument of negation is satisfied
by any object that does not have any other field besides those listed, hence the \qnot: \{ \} schema enforces the presence
of at least one field that does not match the listed patterns. However, the internal description \emph{Invalid additional header field types}  seems to suggest that the intention was rather that of forbidding the presence of the listed fields, which would have been accomplished by the following schema (written by us) - where we wrote \qnot : \{ \qdescr \ldots \},
instead of just \xfalse, in order to show that this second schema can be obtained by the first one by changing the position of 
the negation, and hence its meaning. Of course we do not know whether our interpretation of the \xdescr\ field is correct.
\GG{Somebody volunteers to ask the authors of the schema?}

{\footnotesize
\begin{verbatim}
"type": "object",
"patternProperties": {
    "^date|received_lines|...|subject$": 
         {  "not" : { "description": "Invalid additional header field types" }  }
},
...
\end{verbatim}
}      

Finally, the last 4 schemas present variations of the following schema \refschema{51196}. 

{\small
\begin{verbatim}
"anyOf": [ { "patternProperties": { "^windows-process-ext$": {
                                               "type": "object", ...} } },
           { "patternProperties": { "^windows-service-ext$": {
                                               "type": "object", ...} } },
           { "not": { "patternProperties": {
                            "^windows-service-ext|windows-process-ext$": {
                                    "description": "Invalid custom file extension"
                            }
                      }
             }
           }
]
\end{verbatim}
}

Again,  \qpattProps: \{ \qkey{p} : \{ \qdescr : ... \} \}  is a trivial statement 
since it says: if the instance in an object, then if it has a field that matches the name 
$p$ then its value must match \{ \}, that is, it can be any value. This statement is
satisfied by anything that is not an object, that is an object without a $p$ member,
and that is an object with a $p$ member. Hence, its negation is unsatisfiable.
Since falsehood is a neutral element of disjunction, the entire unsatisfiable
\{ \qnot: \{ \}\} is just redundant.
It is difficult to guess the intention of the authors of this schema, but the impression
is that this redundancy was not their aim. We have also the impressione that the
outermost \xany\ was rather intended to be an \xall, but this would have made the
entire schema unsatisfiable.

If the intention was that of forbidding

 \verb!^windows-service-ext|windows-process-ext$!, 

\noindent this effect can be obtained either with the use of \xaddProps, as previously
discussed, or with a code like the following one, where ``***patternNot...***'' is a pattern that matches the complement of the pattern 

\verb!^windows-service-ext|windows-process-ext$! 

\noindent (observe that \xnot\ is inside 
\xpattProps\ and not outside). Unfortunately, the computation of the complement of 
a pattern is a very complex operation and is not natively supported by {\JS}, hence
the \xaddProps\ approach is the only practical one.

{\small
\begin{verbatim}
"patternProperties": {
      "***patternNot(^windows-service-ext|windows-process-ext$)***":
           { "not": { "description": "Invalid custom file extension" }}}
\end{verbatim}
}

\subsubsection{\xnot.\xaddProps}

The path \xnot.\xaddProps\ appears exactly 11 times in our collections, it always co-occurs with
\xpattProps\ and with a \xfalse\ argument, and these 11 cases are the first 11 cases that we described in the previous section.

\code{
select sibkeys, value, *
from treewithsiblings
where path like '
order by sibnum;
}

\subsubsection{\xnot.\xit}\label{sec:notitems}

The pattern \xnot.\xit, with 126 occurrences in 27 files, is the second most common \xnot-path, which we found 
quite surprising. 

All \xnot.\xit\ schemas have one of the two structures exemplified below  \refschema{88916}, \refschema{21120}.

{\small     
\begin{verbatim}
"not": {"items": {"not": {"$ref": "#/definitions/ewrWithPurpose"}}}
"not": {"items": {"enum": ["ansible",...,"terraform"] } }
\end{verbatim}
}

Hence, all negated schemas with \xit\ are \simple\ schemas that only contain the \xit\ keyword,
where the argument of \xit\ is itself a \simple\ schema, which is always either a \simple\ \xnot\ schema
or a \simple\ \xenum\ schema.
The \xnot.\xit.\xnot\ form is the most common, with 84 occurrences in 22 files, while  \xnot.\xit.\xenum\ occurs 42 times
in 5 files.

The \xit\ assertion is verified by any instance that is not an array, or that is an empty array,
or that is an array where every element satisfies the schema associated with \xit.
Hence, it is only violated by instances that are arrays, and which contain at least one element that
violates the schema. Hence, while \xit\ specifies a univerally quantified properties, \xnot.\xit\ can be used
to specify an existentially qualified property, as does the \xcont\ keyword, according to the following equivalences.

\code{
--instances and files
select key, count(*), count(distinct line), min(line)
from dftree
where path like '
      or path like '
group by rollup(key)

--siblings and children of not.items: no siblings, children either enum or not
SELECT e.p2key, e.pkey, s.sibkeys, json_agg(e.key)
FROM eflattree e
join etreewithsiblings s on (e.pdewey = s.dewey)
--JOIN dfn2 d using (line)
where  not e.added
      and e.p2key = 'not' and e.pkey = 'items'
-- path like '
--and d.description not like '
--and d.description not like '
group by e.p2key, e.line, e.pkey, e.pdewey, s.sibkeys
order by e.line  --, d.description

--following $eref
select value
from edftree
where path like '
}

\qnot: \{ \qit: $S$ \} $\ \Iff\ $
\{\qtype : \qarray, \qcont: \{ \qnot: $S$ \} \} \\[0.7ex]
\qnot: \{ \qcont: $S$ \} $\ \Iff\ $
\{\qtype : \qarray, \qit: \{ \qnot: $S$ \} \} \\[0.7ex]
\{\qtype : \qarray,  \qnot: \{ \qit: $S$ \} \} \\
$\qquad\qquad\ \Iff\ $ \{\qtype : \qarray,  \qcont\{ \qnot: $S$ \} \} \\[0.7ex]
\{\qtype : \qarray,  \qnot: \{ \qcont: $S$ \} \} \\
$\qquad\qquad\ \Iff\ $ \{\qtype : \qarray,  \qit\{ \qnot: $S$ \} \}

Hence, the \xnot.\xit.\xenum\ jargon specifies that the array must contain at least one value that is 
not one of those listed in the argument of \xenum.

The \xnot.\xit.\xnot\ jargon specifies that the instance is an array that
contains at least one value that satisfies $S$, according to the following equivalence.

\qnot: \{ \qit: \{ \qnot: $S$ \} \} $\ \Iff\ $
\{\qtype : \qarray, \qcont: $S$ \} 

The two examples we gave exhaust, with minimal variations, the shapes of the 126 occurrences of
\xnot.\xit. Their extreme homogeneity prompted us to check the nature of the schemas where we found them, 
and we realized that the occurrences are spread in 27 schemas, that belong to just three groups:
(a) ``Annotation'', which is a group of 102 schemas that collectively formalize the \emph{Web Annotation Data Model} 
(https://www.w3.org/TR/annotation-model/), 21 of which use the \xnot.\xit.\xnot\ structure;
(b) ``G-Cloud'', which is a group of 10 files whose title is a variation of
``G-Cloud XX Cloud YY Product Schema'', 5 of which use the \xnot.\xit.\xenum\ structure; (c) ``Other'', which collects all other schemas, where we find one instance of the \xnot.\xit.\xnot\ structure.


\begin{table}\label{tab:notitems}
\center
\small
\begin{tabular}{| l | r| r|  r  |}
\hline
Group   &\xnot.\xit.\xnot &\xnot.\xit.\xenum & Distinct files\\
\hline
Total   &84 &42 &27   \\
Annotation    &    83 & 0  &21    \\
Other &1 & 0 & 1    \\
G-Cloud   & 0 &42 & 5      \\
\hline
\end{tabular}
\caption{Occurrences of \xnot.\xit.\xnot and \xnot.\xit.\xenum}
\end{table}

To sum up, \xnot.\xit\ is used as an alternative way to express \xcont\ constraints.
It appears very often, but this is mostly due to two specific groups that do a massive use of this
jargon. 

This case is quite instructive: it shows that the numbers that we collect are submitted to a bias that depends of the productivity of some specific groups, and this bias can be very strong. This should never be forgotten when these numbers are 
examined.

\code{
--

select key, count(*), count(distinct line), min(line), g.rgroup
from dftree join filetog g using (line)
where path like '
      or path like '
group by rollup(key,g.rgroup)
}

\section{The context around \xnot}\label{sec:contexts}

\subsection{Counting the contexts}

{\JS} is a compositional language, where a negative schema 
\qnot : $S$ may either be used to specify the structure of a property,
or may be bound to a name through the use of \xdefs, or may be 
the argument of another combinator, such as \xone, \xany, \xall, \xite,
\xdeps.

Table \ref{tab:notcontext} shows the context of 774 occurrences of \xnot.
This number is lower than the total of 787 of Table \ref{tab:notargs}, since we did not
count the cases where the context is a user-defined keyword.\footnote{As discussed in Section \ref{sec:frequence},
inside a user-defined keyword we are not even sure that \xnot\ is used as a {\JS} keyword or just as 
a member name} 
The most represented group of contexts is that of boolean operator: if we consider \xone, \xall, \xany,
\xdeps, \xif\ and \xthen\ together, we have a total of 400 occurrences.
The structural operators, such as \xprops\ and \xit, and the top-level contexts, such as \xdefs\ and the root
of the schema, constitute the other half, of 373 contexts.
This is already an informative observation: the frequency of boolean contexts is in the 10-15\% range
for the generic nodes of the schema, but is 52\% for the occurrences of \xnot.

\begin{table}\label{tab:notcontext}
\center
{
\begin{tabular}{| l | r | r|  r | r  |}
\hline
path	& occ.\ of  & freq.\ of \xnot 	&	occ.\ of  & distinct   \\
& the path & $\times$ 10,000 & the context & files \\
\hline
total & 773 & 3 & 624 & 295 \\
properties\xds.\xnot	&	194	&	2	&	187	&	101	\\
oneOf\xarrs.\xnot	&	162	&	11	&	126	&	78	\\
allOf\xarrs.\xnot	&	132	&	130	&	74	&	47	\\
items.\xnot	&	96	&	13	&	96	&	30	\\
anyOf\xarrs.\xnot	&	65	&	20	&	38	&	32	\\
definitions\xds.\xnot	&	43	&	2	&	35	&	35	\\
.\xnot\ (root position)	&	35	&	5	&	35	&	35	\\
dependencies\xds.\xnot	&	24	&	191	&	11	&	8	\\
then.\xnot	&	14	&	385	&	14	&	3	\\
if.\xnot	&	3	&	169	&	3	&	1	\\
patternProperties\xds.\xnot	&	3	&	6	&	3	&	2	\\
additionalProperties.\xnot	&	2	&	3	&	2	&	2	\\
\hline
\end{tabular}
}
\caption{Occurrences of a \key{k}.\xnot\ path for different keywords,
frequency of \xnot\ for every 10,000 nodes in the same position,
and occurrences of the context (e.g., occurrences of \xprops\ followed by .*.\xnot)}
\end{table}

In order to measure the correlation between the different contexts and \xnot, we computed
the ratio between the occurences of \xnot\ in a given position, for example at the end of 
a path \xone\xarrs.\xnot, and the total number of nodes at the end of a path \xone\xarrs.\emph{k},
and we put it in the table. Observe that the global ratio 3/10,000 in the first line 
is not the average of the
other lines, since it includes all the contexts where \xnot\ does not appear at all, that is,
all lines such that, if they were in the table, then their frequence would have a value of 0.
This ratio shows that there is a strong general correlation between \xnot\ and the boolean operators, and it is very strong when the context is \xall, while more generic contexts,
such as \xprops, \xdefs, and the root, do not attract negation in any special way.
The assertion \xdeps\ has a very high frequency of negation, which we
will examine later. We ignore the four contexts after \xdeps\ since none of them
appears in more than 3 distinct schemas.

We now examine all the different contexts in the table up to \xdeps.

\subsection{Boolean contexts}

While the occurrences of \xnot.\key{boolOp} are characterized by an extreme homogeneity of the argument of \key{boolOp},
 (see Section \ref{sec:notany} and following sections), 
the pattern \key{boolOp}.\xnot\ has the opposite behaviour.

Out of 74 instances of \xall.\xnot, 25 are \xnot-homogeneous, that is, all arguments are negated, and
49 are mixed: some of the arguments are positive and some are negative, with a great heterogeneity of situations.

\code{
-- analysing the structure of the argument of xall
select f1.p2dewey, f1.line, count(*) argsOfAllOf,
       jsonb_agg(distinct f2.key) distinctArgsOfAllOf,
       jsonb_agg(f2.key order by f2.key) argsOfAllOf
from flattree f1 join flattree f2 on (f1.p2dewey=f2.p2dewey)
where f1.p2key='allOf' and f1.key ='not'
group by f1.p2dewey, f1.line
order by jsonb_agg(distinct f2.key);

-- counting the total number of arguments of pure xall.not
with pureContexts as
(select distinct f1.p2dewey --, f1.line, count(*), jsonb_agg(distinct f2.key)
from flattree f1 join flattree f2 on (f1.p2dewey=f2.p2dewey)
where f1.p2key='allOf' and f1.key ='not'
group by f1.p2dewey, f1.line
having '["not"]' = jsonb_agg(distinct f2.key) 
)
select count(distinct f.p2dewey) as "allOfFollowedByNot", count(*) as "countAllOfNot"
from flattree f
where f.p2dewey in (select * from pureContexts)

-- looking at the shape of the values
with pureContexts as
(select distinct f1.p2dewey --, f1.line, count(*), jsonb_agg(distinct f2.key)
from eflattree f1 join eflattree f2 on (f1.p2dewey=f2.p2dewey)
where not f1.added and not f2.added and
   f1.p2key='allOf' and f1.key ='not'
group by f1.p2dewey, f1.line
having '["not"]' != jsonb_agg(distinct f2.key) 
--having '["not"]' = jsonb_agg(distinct f2.key) 
)
select f.p2dewey, f.key, f.value
from eflattree f
where f.p2dewey in (select * from pureContexts)
order by f.p2dewey, f.key, f.value

}

After direct analysis of the 49 instances, the impression is that schema designers are typically describing 
an instance that satisfies a schema $S$ but should not present a specific field, or a specific value, or a specific
pattern, hence what they are describing is, essentially, a subtraction, as in the following examples \refschema{89142}, \refschema{4198}: 
an ``amountbase'' but without a \emph{value} field, or a string that is not a ``namespace''.

{\small     
\begin{verbatim}
"allOf":
  [{"$ref": "amountbase"}, {"not": {"required": ["value"]}}]
"allOf":
  [ {"type": "string"} ,  {"not": {"$ref": "#/core/namespace"}} ]
\end{verbatim}
}

While the 49 mixed instances are very different, the 25 pure-not instances all have the following shape \refschema{17072}:
the 25 occurrences have a total of 70 arguments, most of then with shape \xnot : \{ \xreq : [\key{k}] \}.
This is just a different way of expressing the \xnot.\xany\xarrs.\xreq\ pattern that we described in Section \ref{sec:notany}.

{\small     
\begin{verbatim}
"allOf":[{"not": {"required": ["XAngle"]}},
         {"not": {"required": ["YAngle"]}}
         {"not": {"required": ["ZAngle"]}}]
\end{verbatim}
}

The instances of \xone\ followed by some \xnot\ are typically quite complex, as well. 
We have 126 such instances, 
with a total of 320 arguments, 162 of them containing a \xnot. Hence, also in this case, there is a
heavy mixture of negated arguments and positive arguments.
Moreover, half of the 162 arguments containing a \xnot\ are complex.

A typical example of negation in a \xone\ context is the following one \refschema{79022}. 

{\small     
\begin{verbatim}
"type": "string",
"oneOf": [ { "pattern": "^#/components/schemas/"},
           { "not": { "pattern": "^#/" } }
] 
\end{verbatim}
}

The two cases of \xone\ are mutually exclusive; they would actually be both satisfied by any instance
that is not a string, but the external \qtype: \qstr\ assertion excludes these cases.
Hence, the designer is not using \xone\ in order to \emph{force} the instance not to satisfy both branches,
but rather to stress the fact that the two branches are incompatible.
In other terms, here \xone\ could be substituted by \xany\ without changing the semantics,
but \xone\ may convey the designer intention in a way that is clearer.
Hence, this is really a ``not A or B'' pattern, hence an implication: if the string matches
the pattern \QQ\NN\#/\QQ then it must match \QQ\NN\#/components/schemas/\QQ.

Another typical example is the following one.

{\small     
\begin{verbatim}
"dependencies": {
  "construct_type": {
    "oneOf": [
        { "properties": {
            "construct_type": { "enum": ["fusion protein"] },
            "tags": { "minItems": 1  }
            },
          "required": [ "tags" ]
        },
        { "not": {
             "properties": {
                "construct_type": { "enum": ["fusion protein"] }
             }
           }
        }
    ]
}
\end{verbatim}
}

Again, the \xdeps\ context ensures that \xone\ will only be relevant in situations where the instance
is an object with a \verb!construct_type! field, hence the two branches can be regarded as
mutually exclusive, hence the semantics is an implication:
if the value of \verb!construct_type! is \verb!"fusion protein"!, then the \verb!tags! field
is required, and it must satisfy \qminIt: 1. 
We observe that the \xdeps\ assertion of {\JS} allows one to specify that if a field is present
then the instance should satisfy some specific properties. In this case we need to specify that
if a field is present \emph{and} its value satisfies a given type, then the instance should satisfy some specific properties, but this form of dependency is not supported, hence the use of 
disjunction-negation. A slightly simpler way to express this form of dependency would be by using 
the \xif-\xthen\ construct as follows, but this construct has only been introduced with {\VerSeven},
hence its use is very rare.

{\small     
\begin{verbatim}
"if": {
  "required" : ["construct_type"],
  "properties": { "construct_type": { "enum": ["fusion protein"] } }
},
"then": {
  "required" : ["tags"],
  "properties": { "tags": { "minItems": 1 } }
}
\end{verbatim}
}

These two examples do not exhaust the uses of \xone\ followed by \xnot. On the contrary, out of 126 instance of the patter, we counted 41 different structures of use, differing either in the number of arguments of \xone\ or in the set of keywords in this set of arguments.
This variety is quite intersting, but we do not think it is worth to pursue this specific pattern any more, since the different cases are too many, and many of them are quite complex.

As far as \xany\ is concerned, we have a similar situation: out of 38 occurrences of \xany\ where some branch is a negation, 23 mix some arguments that contain negation with arguments that
have no negation, hence encoding an implication. Very few of them are homogeneous, 
like the following one where all three arguments contain the same pair of operators
\xnot-\xreq\ \refschema{4139}: 

{\small     
\begin{verbatim}
"anyOf":[{ "not": {"required": ["image"]},
           "required": ["build"]
         },
         { "not": {
               "anyOf": [ {"required": ["build"]},
                          {"required": ["dockerfile"]}
                        ]
           },
           "required": ["image"]
         },
         { "not": { "required": ["build","image"]},
           "required": ["extends"]
         }
        ]
\end{verbatim}
}

This specification is quite interesting. One notices that the second and the third branch are not
mutually exclusive - they are both satisfied by an instance that has both \emph{image}
and \emph{extends} provided it does not contain neither \emph{build} nor \emph{dockerfile},
and similarly for the first and the third branch.
One also notices that the second branch suggests that \emph{image} and \emph{dockerfile}
are not compatible, but the third branch allows the presence of both, when a field
\emph{extends} is also present.
We conclude that, while \xone\ is quite hard to decode because of the mutual exclusion 
problem, the readability of schemas with \xany\ may be complicated by the lack of a mutual
exclusion property. Finally, this example also shows that real-world schemas 
do present a rich variety of many-levels nesting of boolean operators, that are often simple
to read but sometimes require a bit of effort in order to be decoded.

\subsection{Contexts \xprops\ and \xdefs}

\code{
--count occurrences of simple and not simple: 98 - 96
select count(*) as cocc, count(distinct line) as cfiles, jsonb_agg(line), 
        sibnum, sibkeys, valuelen --
from treewithsiblings
where key = 'not' and path similar to '
group by sibnum, sibkeys, valuelen --
order by cocc desc

--analysiie the structure
select count(*) as cocc, count(distinct line) as cfiles, jsonb_agg(line) 
        sibnum, sibkeys, valuelen --
from treewithsiblings
where key = 'not' and path similar to '
--and sibnum > 1
group by sibnum, sibkeys, valuelen --
order by cocc desc

--counting
select  const::text,
       case when sibnum = 1 then 'simpleSchema' else 'complexSchema' end, 
       case when valuelennew = 1 then 'simpleArg' else 'complexArg' end,
	   count(*) as cocc, 
       count(distinct line) as cfiles, 
	   jsonb_agg(distinct line) files
from treesibargs, ( values ('properties'),('definitions')) as const
where key = 'not' 
        and path similar to concat('
group  by   const,
       cube(
		   case when sibnum = 1 then 'simpleSchema' else 'complexSchema' end, 
       case when valuelennew = 1 then 'simpleArg' else 'complexArg' end )
order by const desc, cocc desc

}

Negation is used at the top level of the definition of a schema for a property in 194 schemas in 101 different files, and in the top level of a schema in the \xdefs\ section in 43 schemas in 35
distinct files (Table \ref{tab:notcontext}).

The 194 occurrences found belows a \xprops\ assertions are uniformly distributed across the range of 
complexity of the schema where \xnot\ appears, and of complexity and operators of the argument
of \xnot.\GG{Clarify}
The 43 occurrences in the \xdefs\ section are, on the contrary, characterized by the fact that 
the schema where \xnot\ appears is typically complex, that is, it combines negation with positive specifications in 39 cases out of 43 as in the following example \refschema{90953}. 
By contrast, in the \xprops\ context the complex schemas are one out of two (96/194), 
and, in the general case, the frequency of complex schemas over the totality of schemas
with \xnot\ is even lower (341 / 787).

{\small     
\begin{verbatim}
"definitions": {
  "common": {
      "type": "object",
      "not": { "description": "cannot have result and error at the same time",
               "required": ["result","error"] },
      "properties": {
          "id": { "type": ["string","integer","null"],
	          "note": [ ... ] }
          },
          "jsonrpc": {"enum": ["2.0"]}
      },
      "required": ["id","jsonrpc"]
  },
  ...
\end{verbatim}
}

The more complex nature of schemas with negation found below \xdefs\ is not surprising:
we have already seen that definitions are mostly used to define complex
object structures once for all, while properties may admit a description that is either simple or
complex, with similar frequency.

\code{
--counting
select count(*) countNegations, count(distinct p2dewey) countDefGroups, count(distinct line)
from flattree
where key = 'not' and p2key = 'definitions'

--visualizing -- this code miss one definition that has `.' inside the definition name
select line, sibnum, sibkeys, valuelen, value
from treewithsiblings
where key = 'not' and path similar to '
}

\subsection{Context \emph{root}}

Negation is used 35 times at the root of the schema, and in 33 cases it respects one of the
following shapes \refschema{64729}, \refschema{13020}, \refschema{91408}: 

{\small     
\begin{verbatim}
"not" : { "anyOf": [ { "required": [ "severity" ] },
                     { "required": [ "action" ] },
                     { "required": [ "usernames" ] },
                     { "required": [ "phone_numbers" ] }
                    ]
}

"not": {"allOf": [{"required": ["site"]},{"required": ["app"]}]}

"not": {"required": ["body","bodyValue"]}
\end{verbatim}
}

The first form lists a set of prohibited member names, while the second (more common)
and the third (quite rare) express mutual exclusion, hence we find, at the root level, the use of
negation that is the most common of all.

The only real surprise with the root context was the following schema \refschema{90941} 
that specifies, in a sense,
that the root implies its own negations, since \{\qdref: "\#"\} is a recursive reference to the 
root of the schema. This form is actually illegal in {\JS}, since it produces an infinite loop.
{\small     
\begin{verbatim}
{ ...
  "not" : {"$ref": "#"},
   ...
}
\end{verbatim}
}
Naturally this is just a ``test'' schema, as testified by its url:\\
\verb!.../json/tests/testData/jsonSchema/highlighting/cycledWithRootRefInNotSchema.json!


\code{
--35 uses
--negation is always not allof required / not anyof required / not required
select *
from flattree d 
where p2key = 'not'
and level = 3
order by pkey;

}

\subsection{Context \xdeps}

\code{
select line, path, sibnum, sibkeys, valuelen, value
from treewithsiblings
where key = 'not' and path similar to '
}

Negations is used after a path \xdeps.$k$ in 24 occasions. In the vast majority of the cases, it is
yet another way to specify field exclusion, as in \refschema{58369}:

{\small     
\begin{verbatim}
"dependencies":
      {"page_action": {"not": {"required": ["browser_action"]}},
       "browser_action": {"not": {"required": ["page_action"]}},
       ...
\end{verbatim}
}

We have found also a couple of situations where negation is used to express more complex
dependencies, as in \refschema{4912}, 
which specifies that, when a member \emph{age\_units} is present,
a member \emph{age} must be present with a value different from \verb!"unknown"!.

{\small     
\begin{verbatim}
"dependencies": {
  "age_units": {"not": { "properties": {"age": {"enum": ["unknown"]}}},
                "required": ["age","life_stage"],
                "comment": "Age units is required if age is specified 
                            as anything but unknown."
        },
        ...
}
\end{verbatim}
}

\hide{
\subsection{Context \xthen}

\code{
select line, path, sibnum, sibkeys, valuelen, value
from treewithsiblings
where key = 'not' and path similar to '
}
}

\hide{

\subsection{Conclusions}

In this section we discovered that boolean operators are a natural context for \xnot, and,
in this situation, it is not rare to find complex nesting of different boolean operators, often
used to describe implication or exclusion between members of an object type, sometimes
linked to the specific value associated with the constant.
Negation is also used to express 
}

\section{Some interesting examples of complex usage}

\subsection{Using negation to express discriminated union}

This piece of code comes from file \refschema{6924}. 
 We have reduced it a little bit
by removing all descriptive fields and we have simplified the type
associated with \verb!"href"! at line 21.

{\footnotesize
\begin{verbatim}     
{
  "not": {
    "anyOf": [ { "properties": { "type": {"enum": ["name"]},
                                 "properties": {
                                      "not": {
                                          "properties": {
                                                "name": { "type": "string",
                                                          "minLength": 1
                                                }
                                           },
                                           "required": ["name"]
                                      }
                                 }
                  }
               },
               { "properties": { "type": {"enum": ["link"]},
                                 "properties": {
                                      "not": {
                                           "type": "object",
                                           "properties": {
                                               "href": {"type": "string"},
                                               "type": {
                                                     "type": "string"
                                               }
                                           },
                                           "required": ["href"]
                                       }
                                 }
                 }
               }
    ]
  },
  "oneOf": [
        { "type": "null" },
        { "type": "object",
          "properties": { "type": { "type": "string",
                                     "minLength": 1      },
                          "properties": { "type": "object" }
           },
           "required": ["type","properties"]
        }
  ]
}
\end{verbatim}
}

This schema is quite hard to read, first of all since it is a bit long and since it uses the name \qtype\ and \qprops\ as property names. Hence, we first rewrite it using the notation of the
tool we developed to reason about {\JS} \cite{}. This notation is a bit more compact and uses
quotes only for user-level names but not for {\JS} keywords, and this should help a bit.

{\footnotesize
\begin{verbatim}     
{ not(anyOf[props["type": const("name"),
                  "props":not({props["name": {type[str],length(1,+inf)};],
                               req["name"]})
                  ;],
            props["type":const("link"),
                  "props":not({type[obj],
                               props["href":type[str],"type":type[str];],
                               req["href"]})
                  ;]
  ]),
  oneOf[
   type[null],
   {type[obj],
    props["type":{type[str],length(1,+inf)},"props":type[obj];],
    req["type","props"]
   }
  ]
}
\end{verbatim}
}

In this notations \xprops\ is abbreviated to \verb!props! (and we did the same with the property name), \xreq\ to \verb!req!, and every operator that may have a variable number of fields, such as \verb!props!, uses [] for its arguments.

We observe immediately the we have a conjunction ``\{\ \xnot , \xone\ \}'' where the \xnot\
part implies that the instance is an object: only an object may violate a \xprops\ constraints.
Hence, any instance that validates the schema will never validate the branch \xtype[\xnull] of
\xone, hence we can substitute \xone\ with the second branch:

{\footnotesize
\begin{verbatim}     
{ not(anyOf[...
  ]),
  {type[obj],
   props["type":{type[str],length(1,+inf)},"props":type[obj];],
   req["type","props"]
  }
}
\end{verbatim}
}

Now, we rewrite the 
\xnot(\xany[\xprops,\xprops])  part

as \xall[\xnot(\xprops),\xnot(\xprops)].

\xall[\xnot(\xprops),\xnot(\xprops)].

{\footnotesize
\begin{verbatim}     
{ allOf[not(props["type": const("name"),
                  "props":not({props["name": {type[str],length(1,+inf)};],
                               req["name"]})
                 ;]),
        not(props["type":const("link"),
                  "props":not({ type[obj],
                                props["href":type[str],"type":type[str];],
                                req["href"]});])
  ],
  { type[obj],
    props["type":{type[str],length(1,+inf)},"props":type[obj];],
    req["type","props"]
  }
}
\end{verbatim}
}

\newcommand{\tprops}{\kw{props}}
\newcommand{\treq}{\kw{req}}
\newcommand{\tnot}{\kw{not}}
\newcommand{\tany}{\kw{anyOf}}
\newcommand{\ttype}{\kw{type}}
\newcommand{\tall}{\kw{allOf}}
\newcommand{\tifThen}{\kw{ifThen}}
\newcommand{\tobject}{\kw{obj}}
\newcommand{\tstr}{\kw{str}}

We have the following equivalences, where \tifThen(A;B) is implication:\\
\tnot(\tprops[\qkey{a}:S, \qkey{b}:T]) $\Iff$ \\
\tnot(\tall [ \tprops[\qkey{a}:S], \tprops[\qkey{b}:T] ]) $\Iff$ \\
\tany [ \tnot(\tprops[\qkey{a}:S)], \tnot(\tprops[\qkey{b}:T] ])) $\Iff$ \\
\tifThen( \tprops[\qkey{a}:S], \tnot(\tprops[\qkey{b}:T]) ]) $\Iff$ \\
\tifThen( \tprops[\qkey{a}:S], \{\ttype[\tobject],\treq[\qkey{b}],\tprops[\qkey{b}:\tnot(T)] \} )

We use it in order to rewrite the two \tnot(\tprops) in the expression above.
In the process, we transform the nestes conjunction ``\{ \xall[A,B],C \}'' into a 
flat conjunction ``\{ A', B', C \}''. 

{\footnotesize
\begin{verbatim}     
{ ifThen(props["type": const("name");];
         {type[obj],
          req["props"],
          props["props":not(not({props["name": {type[str],length(1,+inf)};],
                                 req["name"]}))
          ;]}
  ),
  ifThen(props["type":const("link");];
         {type[obj],
          req["props"],
          props["props":not(not({type[obj],
                                 props["href":type[str],"type":type[str];],
                                 req["href"]})
                        )
               ;] }
  ),
  type[obj],
  props["type":{type[str],length(1,+inf)},"props":type[obj];],
  req["type","props"]
}
\end{verbatim}
}

Now, the statements \ttype[\tobject] and \treq[\qkw{props}] in the \xthen\ branch of the 
first \tifThen\ are redundant and can be removed, since they are implied by the top-level
\ttype[\tobject] and \treq[\qkw{type},\qkw{props}] found at the end of the schema, 
and the same for the second \tifThen.
And we also eliminate the double negation in front of the schema for the \verb!"props"!
property.


{\footnotesize
\begin{verbatim}     
{ ifThen(props["type": const("name");];
         props["props":{ props["name": {type[str],length(1,+inf)};],
                         req["name"]}
         ;]
  ),
  ifThen(props["type":const("link");];
         props["props": { type[obj],
                          props["href":type[str],"type":type[str];],
                          req["href"] }
         ;]
  ),
  type[obj],
  props["type":{type[str],length(1,+inf)},"props":type[obj];],
  req["type","props"]
}
\end{verbatim}
}

Now the specification is clear: the instance is an object, and it must contain at least fields
\verb!"type"!, whose value is a non-empty string, and \verb!"props"!, whose value is
an object.
If the value of \verb!"type"! is \verb!"name"!, then the value of  \verb!"props"! 
must contain a \verb!"name"! member whose value is a non-empty string. 
If the value of \verb!"link"! is \verb!"name"!, then the value of  \verb!"props"! 
must contain an \verb!"href"! member whose value is a string and, if it contains
a  nested \verb!"type"! member, its value must be a string.

This example is quite interesting, since it shows
three different things:

\begin{enumerate}
\item In order to understand a schema that contains negation, one absolutely
needs a reasonable notation, a clear semantics, and some rewrite rules 
\item Programmers are really using negation with some sophistication.
Here, in a nutshell, the designer has used: 
{\footnotesize
\begin{verbatim}    
not(props["discr": const("caseOne"),
          "body": not({ req["oneField"], props["oneField": { T };] } )
    ;])
\end{verbatim}
}
in order to say that, when \verb!"discr":"caseOne"!, then the member 
\verb!"body"! must contain the \verb!"oneField"! member, with type $T$.
This is not an
unreasonable way to express that specification, but is very difficult to decode.
\item It also shows that {\JS} would need a syntax to express value-to-type
dependencies or, at least, discriminated unions. That would have solved
this example in a totally natural way. 
\end{enumerate}

\subsection{Field exclusion}

In file \refschema{58639} 
we have found a use of negation that is quite surpriing.

{\small
\begin{verbatim}     
"dependencies": {
  "name": { "not": { "required": ["name"] } },
  "icons": { "not": { "required": ["icons"] } },
  "popup": { "not": { "required": ["popup"] } }
}
\end{verbatim}     
}

At first sight, this may look like a contradiction: if \verb!"name"! is present, then it must be absent.
Actually, the meaning is quite simple: the three listed fields cannot appear in the instance. This could also be expressed in many simpler ways, using any of the four following forms. All these expressions are equivalent if the context implies that the instance is an object, otherwise the third and fourth
ones are stronger, since they force the instance to actually be an object.

{\small
\begin{verbatim}     
"dependencies": { "name": { "not": {  } }, 
                  "icons": { "not": {  } }, 
                  "popup": { "not": {  } } }

"properties": { "name": { "not": {  } }, 
                "icons": { "not": {  } }, 
                "popup": { "not": {  } } }

"allOf": [ {"not": { "required": ["name"] }}, 
           {"not": { "required": ["icons"] }}, 
           {"not": { "required": ["popup"] }} ]

"not": { "anyOf": [ { "required": ["name"] }, 
                    { "required": ["icons"] },  
                    { "required": ["popup"] } ] }
\end{verbatim}     
}

\subsection{Discriminated unions, again}

The following code comes from \refschema{90970} 
and is quite puzzling because of the
the use of \xone. 

{\small
\begin{verbatim}     
{ "type" : "object",
  "oneOf": [
     { "properties": {"when": {"enum": ["delayed"]}},
       "required": ["when","start_in"]
     },
     { "properties": {"when": { "not": {"enum": ["delayed"]}}}
     }
  ]
}
\end{verbatim}     
}

We first verify that the two branches of \xone\ are mutually exclusive.
If \verb!"when"! is absent, then only the second branch holds. If it is present,
then is associated to complementary types in the two branches. Hence, \xone\ is
really \xany, hence it can be rewritten as an implication: if the second branch is false then
the first one holds.
{\small
\begin{verbatim}     
{ "type" : "object",
  "if": { "not" : { "properties": {"when": { "not": {"enum": ["delayed"]}}}
                  }
        }
  "then" :  { "properties": {"when": {"enum": ["delayed"]}},
              "required": ["when","start_in"]
     }
}
\end{verbatim}     
}

We rewrite the negation in the usual way.

{\small
\begin{verbatim}     
{
  "type": "object",
  "if": { "required": ["when"],
          "properties": {"when": {"enum": ["delayed"]}}
  },
  "then": { "properties": {"when": {"enum": ["delayed"]}},
            "required": ["when", "start_in"]
  }
}
\end{verbatim}     
}

And now the specification is clear: if \verb!"when"! has the value \verb!"delayed"!,
then \verb!"start_in"! is required. If \verb!"when"! is missing, or has any other value, then
the specification is satisfied.

\subsection{Discriminated unions, again}\label{sec:implication}

File \refschema{37789}. 

\subsection{Negation and recursions}

The interaction between negation and recursion is very problematic. Prolog and Datalog research produced
a great wealth of techniques in order to define the semantics of this difficult combination, while the
$\mu$-calculus, which is the logic languages with greater similarity, just forbids recursion under negation.

JSON Schema allows recursive negation, provided that it does not create an infinite loop in the checking
process. The no-loops rule is actually restraining the use of recursion, and is unrelated with the presence of
negation in the loop. For example, the following two schemas are both ok --- \qdref : \qkw{\#} is a recursive
refernce to the current schema. 

{\small
\begin{verbatim}     
{
  "type": "object",
  "properties": {"foo": {"$ref": "#"}}
}

{
  "type": "object",
  "properties": {"foo": {"not": {"$ref": "#"}}}
}
\end{verbatim}     
}

While the following two schemas are not legal, since they would both cause an infinite loop.

{\small
\begin{verbatim}     
{
  "$ref": "#"
}

{
  "not": {"$ref": "#"}
}
\end{verbatim}     
}

Recursion may be direct, with a schema referring to itself, as in the two examples above,
or indirect, with two different definitions referring the one to the other, as in the following example.

{\small
\begin{verbatim}     
{
  "$ref": "#/definitions/d1",
  "definitions" : { 
  		"d1" : {"$ref": "#/definitions/d2"},
  		"d2" : { "not": { "$ref": "#/definitions/d1"} }
  }
}
\end{verbatim}     
}

We checked whether we could find examples of recursion with negation, and indeed we found one example
of direct recursion with usage of \xnot.
Consider the following snippet \refschema{1520}, 
of which we found 4 versions, all very similar.

{\small
\begin{verbatim}     
{ "oneOf": [ { "$ref": "#/definitions/MSC.Section-common" },
	     ,...
  ],
  "definitions": {
    "MSC.Section-common": {
      "type": "object",
      "properties": {
        ...,        
        "elements": {
          "type": "array",
          "items": {
            "allOf": [ { "$ref": "#" },
                       { "not":
                           { "$ref": "#/definitions/MSC.Section-common" }
                       }
                     ]
          }
        }
      }
    }
  }
}
\end{verbatim}     
}

It specifies that a \emph{Control} (this is the name of the schema) may be a \emph{Section}, which is an object with an array of elements
each of them being itself a \emph{Control} (\qdref : \qkey{\#}) but not a \emph{Section}.

This is another example \refschema{12177}, 
 this time of indirect recursion, of which we found two copies.

{\small
\begin{verbatim}     
{ "type": "object",
  "required": [ ..., "properties" ],
  "properties": {
      ...,
      "properties": {
        "type": "object",
        "additionalProperties": {
          "$ref": "#/definitions/propertyObject"
        }
      }
  },
    "definitions": {
      "propertyObject": {
        "allOf": [
          { "not": { "$ref": "#" } },
          { "$ref": "dataTypeBase.json#" }
        ]
      }
  }
}
\end{verbatim}     
}

Here, we have a \emph{modelObject} (the name of the schema file)
with an mandatory \emph{properties} member, whose members
may have any name, since their schema is specified using \xaddProps, but whose schema must obey 
\emph{propertyObject}. A \emph{propertyObject} must satisfy \emph{dataTypeBase} but must not be a 
\emph{modelObject} itself.

Finally we found a file \refschema{90941} 
which exhibits the forbidded \xnot.\xdref : \#\ pattern, but the name of the
file indicates that this is just a test.  

{\small
\begin{verbatim}     
{
  ...,
  "not": {"$ref": "#"}
}
\end{verbatim}     
}

\code{
select anc.dewey, des.key, des.value, des.path
       --, case when danc.value = des.value then 'yes' else 'no' end as equal
from dftree danc, edftree anc, edftree des
where  des.key = '$ref'
       and des.dewey similar to concat(anc.dewey,'.
       and des.path similar to '
                 and anc.key = '$eref'
                 and danc.dewey = anc.dewey
                 and danc.value = des.value
                 
create table depends as
(select distinct
       --danc.dewey, des.dewey, 
	   danc.value as var ,
       des.value as dependsOn, des.path
       --, case when danc.value = des.value then 'yes' else 'no' end as equal
from dftree danc, edftree anc, edftree des
where  des.key = '$ref'
       and des.dewey similar to concat(anc.dewey,'.
       and des.path similar to '
                 and anc.key = '$eref'
				 and danc.key = '$ref'
                 and danc.dewey = anc.dewey
                 --and danc.value = des.value
)

}

\section{Conclusions}


The lessons drawn from the analysis of negation usage are numerous. 

We first learned that negation is used in many different ways, many more than we would have imagined,
some of which are extremely creative, although some typical cases are very common: field exclusion,
first of all, value exclusion from sets of strings, and field mutual exclusion. 
Field exclusion is so common that one may imagine to add an ad-hoc operator to {\JS}.

Another general lesson that we learned is that {\JS} can be extremely tricky to decode, and the fact that 
JSON notation is used does not help at all.

We observed that negation is also often used in order to express, in a cumbersome way, the \emph{discriminated
union} pattern, where the value of one field determines the presence/absence and the type of the others.
This fact may also trigger some reflections about schema design. 

We also observed the use of negation in order to force the presence of a member whose name satisfies a pattern.
We already noticed that {\JS} is, in some sense, missing this operator, 
and we found here a 
confirmation of that.

Finally, we think that our effort proved that automatic tools to assist the task of decoding {\JS} specification could be
quite useful, which is a motivation to expand our research in that direction.

\nop{
\smallskip
\noindent
\begin{small}
\textit{Acknowledgments.}
This contribution was partly funded by {\em Deutsche For\-schungsgemein\-schaft}\/ (DFG, German Research Foundation) grant \#385808805. 
The schemas were retrieved using Google BigQuery, supported by Google Cloud.
We thank Edson Lucas and Thomas Kirz 
for the charts, and Thomas Pilz for preparing the schema repository.
\end{small}
}

%
%
\bibliographystyle{plain}
\bibliography{references}

\end{document}